\documentclass[final]{article}

\usepackage{graphicx}
\usepackage{dsfont}
\usepackage{amsfonts}
\usepackage{amsmath}
\usepackage{booktabs}

\newtheorem{lem}{Lemma}

\newtheorem{notations}{Notations}
\newtheorem{cor}{Corollary}
\newtheorem{proposition}{Proposition}

\newcommand{\Id}{\mathbb{I}}

\newcommand{\lap}[1]{\widehat {#1}}

\newcommand{\lag}{\tau}
\newcommand{\Ma}{Y}

\newcommand{\lPsi}{\lap\Psi}
\newcommand{\lPhi}{\lap\Phi}

\newcommand{\E}{E}

\newcommand{\lDelta}{\lap D}
\newcommand{\myDelta}{D}

\newcommand{\lCtt}{\lap E^T}
\newcommand{\lCtn}{\lap E^I}
\newcommand{\lCnt}{\lap E^F}
\newcommand{\lCnn}{\lap E^N}

\newcommand{\lphi}{\lap\phi}

\newcommand{\Lambdan}{{\Lambda^N}}

\newcommand{\Phinn}{\Phi^N}
\newcommand{\lPhinn}{{\lap \Phi}^N}
\newcommand{\phinn}{{\phi^{N}}}
\newcommand{\lphinn}{\lap \phi^N}
\newcommand{\lphinnstar}{\lap \phi^{N*}}
\newcommand{\phinns}{\phi^{N,s}}
\newcommand{\phinnc}{\phi^{N,c}}
\newcommand{\lphinns}{\lap\phi^{N,s}}
\newcommand{\lphinnc}{\lap\phi^{N,c}}

\newcommand{\lambdan}{{\lambda^N}}

\newcommand{\Phitn}{\Phi^{I}}

\newcommand{\lPhitn}{{\lap \Phi}^I}
\newcommand{\lPhitnstar}{{\lap \Phi}^{I*}}
\newcommand{\phitn}{\phi^{I}}
\newcommand{\lphitn}{\lap \phi^I}
\newcommand{\phitns}{\phi^{I,s}}
\newcommand{\phitnc}{\phi^{I,c}}

\newcommand{\phitts}{\phi^{T,s}}
\newcommand{\phittc}{\phi^{T,c}}
\newcommand{\lphitts}{\lap\phi^{T,s}}
\newcommand{\lphittc}{\lap\phi^{T,c}}

\newcommand{\lphitns}{\lap\phi^{I,s}}
\newcommand{\lphitnc}{\lap\phi^{I,c}}
\newcommand{\phint}{\phi^{F}}
\newcommand{\lphint}{\lap \phi^F}
\newcommand{\lphintstar}{\lap \phi^{F*}}
\newcommand{\Phint}{\Phi^{F}}
\newcommand{\phints}{\phi^{F,s}}
\newcommand{\phintc}{\phi^{F,c}}
\newcommand{\lphints}{\lap\phi^{F,s}}
\newcommand{\lphintc}{\lap\phi^{F,c}}
\newcommand{\lPhint}{\lap\Phi^F}
\newcommand{\lPhintstar}{\lap\Phi^{F*}}

\newcommand{\Thetan}{{\Theta^{I}}}
\newcommand{\Thetat}{{\Theta^{T}}}

\newcommand{\lthetat}{\lap\theta^{T}}

\newcommand{\Lambdat}{{\Lambda^T}}

\newcommand{\Phitt}{\Phi^T}
\newcommand{\lPhitt}{{\lap \Phi}^T}

\newcommand{\phitt}{{\phi^T}}
\newcommand{\lphitt}{\lap \phi^T}
\newcommand{\lambdat}{{\lambda^T}}

\newcommand{\Mu}{{M}}

\begin{document}

\title{Hawkes model for price and trades high-frequency dynamics}

\author{Emmanuel Bacry\thanks{CMAP, UMR 7641 CNRS, Ecole Polytechnique, 91128 Palaiseau, France
({\tt emmanuel.bacry@polytechnique.fr})} \and
Jean-Fran\c{c}ois Muzy\thanks{
SPE, UMR 6134 CNRS, Universit\'e de Corse, 20250 Corte, France
({\tt muzy@univ-corse.fr})}
}

\maketitle
\begin{abstract}
We introduce a multivariate Hawkes process that
accounts for the dynamics of market prices through the impact of market order arrivals at microstructural level.
Our model is a point process mainly characterized by 4 kernels associated with respectively
the trade arrival self-excitation, the price changes mean reversion
the impact of trade arrivals on price variations and the feedback
of price changes on trading activity. It allows one
to account for both stylized facts of market prices microstructure
(including
random time arrival of price moves,  discrete price grid, high frequency mean reversion, correlation
functions behavior at various time scales) and the stylized facts of
market impact (mainly the concave-square-root-like/relaxation characteristic shape of the market impact of a meta-order).
Moreover, it allows one
to estimate the entire market impact profile from anonymous market data.
We show that
these kernels can be estimated from the empirical
conditional mean intensities. We provide numerical examples, application to real data and
comparisons to former approaches.
\end{abstract}



\pagestyle{myheadings}
\thispagestyle{plain}

\section{Introduction}
Market impact modeling (i.e, the influence of market orders on forthcoming prices) is a longstanding
problem in market microstructure literature and is obviously of great interest for practitioners (see e.g., \cite{bouchaud2} for a recent review).
Even if there are various ways to define and estimate the impact
associated with an order or a meta-order\footnote{One usually refers to a meta-order as a set of orders corresponding to a fragmentation of a single large volume order in several successive smaller executions}, a large number of empirical results
have been obtained recently.
The theory of market price formation and the relationship between the order flow and price changes
has made significant progress during the last decade \cite{BFL09}. Many empirical studies have provided evidence
that the price impact  has, to many respects, some universal properties and is the main source of price variations. This corroborates the picture of an ``endogenous'' nature of price fluctuations that contrasts with the classical scenario according to which an ``exogenous'' flow of information leads
the prices towards a ``true'' fondamental value \cite{BFL09}.

We will not review in details all these results but simply recall
the point of view of Bouchaud et al \cite{bouchaud2}.
These authors propose a model of price fluctuations by generalizing Kyle's pioneering approach \cite{Kyle} according to which
the price is written (up to a noise term) as the result of the impact of all trades.
If $n$ stands for the trading time, Bouchaud et al. model is written as follows \cite{bouchaud1,BFL09}:
\begin{equation}
\label{eqbouchaud}
   p_n = \sum_{j \leq n} G(n-j)\xi_j  + \eta_j
\end{equation}
where $\eta_j$ is a white noise while $\xi_j = \varepsilon_j f(v_j)$
with $\varepsilon_j=+1$ (resp. $\varepsilon_j = -1$) if a trade occurs at the best ask (resp. at best bid)
and $f(v_j)$ describes the volume dependence of a single trade impact. The function $G(n-j)$
accounts for the temporal dependence of a market order impact.

Even if models like \eqref{eqbouchaud} represent a real breakthrough in the understanding of price dynamics
they have many drawbacks. First, the nature and the status of the noise $\eta_j$ is not well defined.
More importantly, these models involve discrete events (through the trading or event time) only defined
at a microstructural level though they intend to represent some coarse version of prices (indeed, the price
$p_n$ in the previous equation can take arbitrary continuous values, moreover, calibrating its volatility at any scale involves some
additional parameter, ...).
Moreover, these models cannot account for real time (i.e ``physical time'') dynamics
or real time aggregation properties of price fluctuations.
In that respect, they are not that easy to be used in real high frequency applications such as optimized execution.
To make it short, though being defined at finest time scales, they cannot account for the main microstructure
properties of price variations related to their discrete nature: prices live on tick grids and jump
at discrete random times.

Our aim in this paper is mainly to define {\em continuous time} version of the market impact price model
discussed previously.
For that purpose, point processes \cite{daley2003introduction} provide a
natural framework. Let us not that point processes have been involved
in many studies in high frequency finance from the famous `zero-intelligence'' order-book
models \cite{Farmer2002StatAuction,Farmer2005ZI} to models for trade
\cite{Hewlett2006Clustering,BH} or book events \cite{Ioane2011Hawkes,Embrechts2011Hawkes} irregular arrivals.
In a recent series of papers \cite{BDHM2010,BDHM2011,BaMuDaEPJB}, we have shown
that self-excited point (Hawkes) processes can be pertinent
to model the microstructure of the price and in particular to reproduce the shape of the signature plot and the Epps effect.
Our goal is to extend this framework in order to account for the market impact of market orders.
In that respect, the main ideas proposed in refs. \cite{bouchaud1,gatheral,FarmerLillo04} can
be reconsidered within the more realistic framework of point processes where correlations
and impact are interpreted as cross and self excitations mechanisms acting on the conditional intensities of Poisson processes.
This allows us to make a step towards the definition of a faithful model of price microstructure that accounts for most recent
empirical findings on the liquidity dynamical properties and to uncover new features.

The paper is organized as follows. In Section \ref{sec:defmodel} we
show how market order impact can be naturally accounted within the class of multivariate Hawkes processes.
Our main model for price microstructure and market impact is presented and
its stability is studied. Numerical simulations of the model are presented.
In Section \ref{covmatrix}, the microstructure of price and market order flows are studied through the covariance matrix of the process.
Our results are illustrated on numerical simulations.
An extension of the model that accounts for labeled agents is defined in Section \ref{M11} and results on market impact are presented in Section \ref{sec:mi} including an explanation on how the newly defined framework allows one to estimate the market impact profile from non labeled data.
Section \ref{sec:estimation} shows how kernels defining the dynamics of trade occurrences
and price variations can be nonparametrically estimated.
All the theoretical results obtained in the previous Sections are illustrated in Section \ref{sec:empirical}
when applied on various high frequency future data. It allows one to reveal the different dynamics involved in price movements, market order flows  and market impact. Intraday seasonalities are shown to be taken care in a particularly simple way.
We also discuss, on a semi-qualitative ground, how the market efficiency can be compatible with the observed
long-range correlation in supply and demand without any parameter adjustment.
Conclusion and prospects are reported in Section \ref{conclusion} while technical
computations are provided in Appendices.

\section{Hawkes based model for market micro\-structure}
\label{sec:defmodel}

\subsection{Definition of the model}

As recalled in the introduction, in order to define a realistic microstructure price
model while accounting for the impact of market orders, the framework of multivariate self-excited
point processes is well suited.
A natural approach is to associate a point process to each set of events
one wants to describe. We choose to consider all market order events
and all mid-price change events.

Let us point out that we will not take into account the volumes associated to each market orders.
Though this can be basically done within the framework
of {\em marked} the point processes, it would necessitate cumbersome notations and make the estimation much more
difficult. This issue be discussed briefly in Section \ref{conclusion} and addressed in a forthcoming work.

\subsubsection{Market orders and price changes as a 4 dimensional point process}
The arrivals of the  market orders are represented by a two dimensional counting process
\begin{equation}
T_t =  \left(\begin{array}{c} T^-_t  \\ T^+_t  \end{array} \right)
\end{equation}
representing cumulated number of market orders  arrived before time $t$ at the best ask
($T^+_t$) and at the best bid ($T^-_t$).
Each time there is a market order, either $T^+$ or $T^-$ jumps up by 1.
We suppose that the trade process\footnote{In the following, a "trade" will refer to the execution of a given market order (which might involve several counterparts). Thus the process $T$ will be referred indifferently to as the market order arrival process or the trade arrival process.
}
$T$ is a counting process that is fully defined by
\begin{itemize}
\item $\lambdat_t$ the conditional intensity vector of the process $T_t$ at $t$.
\end{itemize}

The price is represented in the same way.
Let $X_t$ represents a proxy of the price at high-frequency (e.g., mid-price). As in refs. \cite{BDHM2010,BDHM2011,BaMuDaEPJB}
we write
\begin{equation}
\label{eq:X}
X_t = N^+_t-N^-_t
\end{equation}
where $N^+_t$ (resp. $N^-_t$) represents the number of upward (resp. downward) price jumps at time $t$ \footnote{Let us point out that, in this model, we do not take into account the size of the upward or downward jumps of the price. We just take into account the direction of the price move, $+1$ (reps. $-1$) for any upward (reps. downward) jumps}.
Thus, each time  the price goes up (resp. down), $N^+$ (resp. $N^-$) jumps up by 1.
We set
\begin{equation}
N_t =  \left(\begin{array}{c} N^-_t  \\ N^+_t  \end{array} \right).
\end{equation}
As for the trade process, $N_t$ is fully defined by
\begin{itemize}
\item $\lambdan_t$ the conditional intensity vector of the process $N_t$ at $t$.
\end{itemize}
The 4 dimensional counting process is then naturally defined as
\begin{equation}
  P_t =
 \left(\begin{array}{c} T_t  \\ N_t  \end{array} \right)
  =
   \left( \begin{array}{c}
    T^-_t \\
    T^+_t \\
    N^-_t \\
    N^+_t
    \end{array} \right) \; .
\end{equation}
as well as its associated conditional intensity vector
\begin{equation}
  \lambda_t =
 \left(\begin{array}{c} \lambdat_t  \\ \lambdan_t  \end{array} \right)
  =
   \left( \begin{array}{c}
    \lambdat^-_t \\
    \lambdat^+_t \\
    \lambdan^-_t \\
    \lambdan^+_t
    \end{array} \right) \; .
\end{equation}


\subsubsection{The Model}
\label{M1}
Basically, the model consists in considering that the 4-dimensional counting process $P_t$ is a Hawkes process \cite{Hawkes1971-1,Hawkes1971-2}. The structure of a Hawkes process allows one to take into account the influence of any component in $P_t$ on any component of $\lambda_t$.
In its general form the model is represented by the following equation\footnote{Let us point out that Section \ref{M11}  will introduce a generalization of this model including labeled trades.}
\begin{equation}
\label{Hawkes4d}
\lambda_t = \Mu + \Phi \star dP_t .
\end{equation}
where $\Phi_t$ is a $4 \times 4$ matrix whose elements are {\em causal} positive functions (by {\em causal} we mean functions
supported by $\mathds{R}^+$). Moreover, we used the ``matrix convolution" notation,
\begin{equation}
\nonumber
B \star dP_t=\int_{\mathds{R}}B_{t-s} dP_s,
\end{equation}
where $B_{t-s} dP_s$ refers to the regular matrix product. $\Mu$ accounts for the exogenous intensity of the trades, it has the form
\begin{equation}
\Mu =  \left(\begin{array}{c}
\mu \\
\mu \\
0 \\
0
 \end{array} \right)
\end{equation}
since, by symmetry we assume that the exogenous intensity of $T^+$ and $T^-$ are equal while
mid-price jumps are only caused by the endogenous dynamics.

The matrix  $\Phi$ or any of its sub-matrices (or element) are often referred to as Hawkes kernels. Each element describes the influence of a component over another component. Thus, it is natural to decompose the $4\times 4$  kernel $\Phi_t$ in four $2\times 2$ matrices in the following way
\begin{equation}
  \Phi_t = \left(\begin{array}{cc}
\Phitt_t & \Phint_t \\
\Phitn_t & \Phinn_t
 \end{array} \right)
\end{equation}
where
\begin{itemize}
\item $\Phitt$ (influence of $T$ on $\lambdat$) : accounts for the trade correlations (e.g., splitting, herding, \ldots).
\item $\Phitn$ (influence of $T$ on $\lambdan$) : accounts for the impact of a single trade on the price
\item $\Phinn$ (influence of $N$ on $\lambdan$) : accounts for the influence of past changes in price on future changes in price (due to cancel and limit orders only, since
changes in price due to market orders are explicitly
 taken into account by $\Phitn$)
\item $\Phint$ (influence of $N$ on $\lambdat$) : accounts for {\em feedback} influence of the price moves on the trades.
\end{itemize}
If we account for the obvious symmetries between bid-ask sides for trades and up-down directions for price jumps these
matrices are naturally written as:
\begin{equation}
\label{kernel1}
\Phitt_t =  \left(\begin{array}{cc}
\phitts_t & \phittc_t \\
\phittc_t & \phitts_t
 \end{array} \right), \; \; \;
\Phitn_t =  \left(\begin{array}{cc}
\phitns_t & \phitnc_t \\
\phitnc_t & \phitns_t
 \end{array} \right)
\end{equation}
and
\begin{equation}
\label{kernel2}
\Phinn_t =  \left(\begin{array}{cc}
\phinns_t & \phinnc_t \\
\phinnc_t & \phinns_t
 \end{array} \right), \; \; \;
\Phint_t =  \left(\begin{array}{cc}
\phints_t & \phintc_t \\
\phintc_t & \phints_t
 \end{array} \right)
\end{equation}
where all $\phi_t^{?,?}$ are causal functions and the upperscripts $s$ and $c$ stand for ``self'' and ``cross'' influences
of the Poisson rates (we use the same convention which was initially introduced  in  \cite{BDHM2010} for  $\phinn$).
Thus, for instance, on the one hand,
 $\phitns$ accounts for the influence of the past buying (resp. selling) market orders on the intensity of the future upward (reps. downward) price jumps. On the other hand,
 $\phitnc$ quantifies the influence of the past buying (resp. selling) market orders on the intensity of the future downward (resp. upward) price jumps.

\vskip .2cm
\noindent
{\bf Remark 1 : } All these $2\times 2$  matrices commute since they diagonalize in the same basis (independently of $t$).
Their eigenvalues are the sum (resp. the difference) of the self term with the cross term. This property will be used all along the paper. Most of the
computations will be made after diagonalizing all the matrices.


\subsubsection{The Impulsive impact kernel model or how to deal with simultaneous jumps in the price and trade processes}
\label{impulsive}
It is important to point out that a buying market order that eats up the whole volume sitting at best ask results in an ``instantaneous"  change in the mid-price. From our model point of view, it would mean that $T$ and $N$ have simultaneous jumps with a non zero probability. It is clearly not allowed as is by the model. However, from a numerical point of view, this can be simulated by just considering that the jump in the price takes place within a very small time interval
(e.g., of width 1ms which is the resolution level of our data) after the market order has arrived. There is no ambiguity in the ``direction" of the causality :
it is a market order that makes the price change and not the other way around.
This would result in an impact kernel $\phitns$  which is ``impulsive", i.e., localized around 0,
actually close to a Dirac distribution $\delta_t$.

From a practical numerical point of view, choosing $\phitns$ to be a Dirac distribution is fairly easy. It basically amounts to considering that it is a positive function of given $L^1$ norm $I$ and with a support $\Delta t$ of the order of a few milliseconds. Let us point out that it means that, the price increment between the moment of the trade and $\Delta t$  milliseconds afterwards, follows a Poisson law whose parameter is $I$. This actually allows price jumps (spread over a few milliseconds) of several ticks (greater than $1$).

Of course,
from a mathematical point of view, this is not that simple. The limit $\Delta t \rightarrow 0$ has to be defined properly. This will be rigorously defined and extensively discussed in a future work and is out of the scope of this paper. The "practical" approach described above and the fact that we can formally replace, in all the computations,  $\phitns$ by a $I\delta_t$
is far enough for our purpose.

It is clear that we expect to find an impulsive component in $\phitns$ when estimating on real data. Though, a priori, we do expect also a non singular component that could have a large support (e.g., when the marker order eats up only a part of the volume sitting at best ask), we will see in estimations that most of the energy of $\phitns$ is localized around 0. Moreover, we will find that $\phitnc$ is close to 0.

All these remarks will lead us to study a particularly interesting case of the previously defined model for which
\begin{itemize}
\item $\phitns = I\delta_t$
and  $\phitnc = 0$.
\end{itemize}
Consequently
\begin{itemize}
\item $\phitn = I\delta_t\Id$, where $\Id$ refers to the identity matrix.
\end{itemize}
This model will be referred in the following as the Impulsive Impact Kernel model.

Before moving on and study the conditions for our model to be well defined,
we need to introduce some notations that will be used all along the paper.

\subsubsection{Notations}
Let us introduce the following notations:
\begin{notations}
If $f_t$ is a function $\lap f_z$ refers to the Laplace transform of this function, i.e.,
\[
\lap f_z = \int e^{-izt}f_t dt.
\]
$\delta_t$ refers to the Dirac distribution, consequently
\[
\lap \delta_z = 1.
\]
Moreover, we will use the convenient convention  (which holds in the Laplace domain)~:~$\delta_t  \star \delta_t  = \delta_t $.  \\
The $L^1$ norm of $f$ is referred to as:
\[
||f|| = \int |f_t| dt.
\]
Thus
\[
\mbox{if~~~} \forall t,~f_t \ge 0  \mbox{~~then~~} ||f|| = \int f_t dt  = \lap f_0.
\]
We extend these notations to matrix of functions. Thus, if $F_t$ is a matrix whose element are functions of $t$, let $\hat F_z$ denote the matrix whose elements are the Laplace transform of the elements of $F_t$. Following this notation, we note
\begin{equation}
\label{lphi}
\lPhi_z =  \left(\begin{array}{cc}
\lPhitt_z & \lPhint_z \\
\lPhitn_z & \lPhinn_z
 \end{array} \right).
\end{equation}
\end{notations}

\begin{notations} If $M$ is a matrix,
$M^*$ refers to the matrix $M$ whose each element  has been replaced by its conjugate and
$M^\dagger$ refers to the hermitian conjugate matrix of $M$.
\end{notations}

\begin{notations} Whenever $\lambda_t$ is a stationary process, we will use the notation
\[
\Lambda = \E(\lambda_t) =  \left(\begin{array}{c}
\Lambdat \\
\Lambdat \\
\Lambdan \\
\Lambdan
 \end{array} \right)
\]
where
\[
\Lambdat = \E(\lambdat^{+}_t) = \E(\lambdat^-_t) ~~~ \mbox{and}~~~ \Lambdan = \E(\lambdan^{+}_t) =  \E(\lambdan^{-}_t).
\]
Let us point out that the fact that the $\pm$ mean intensities are equal is due to the symmetries of the kernels in \eqref{kernel1} and \eqref{kernel2}.
\end{notations}

\begin{notations}
\label{not:delta} We define the {\em kernel's imbalance} :
\begin{itemize}
\item $\Delta \phitt = \phitts-\phittc$  and $\Delta \lphitt = \lphitts-\lphittc$
\item $\Delta \phitn = \phitns-\phitnc$ and $\Delta \lphitn = \lphitns-\lphitnc$
\item $\Delta \phint = \phints-\phintc$ and $\Delta \lphint = \lphints-\lphintc$
\item $\Delta \phinn = \phinns-\phinnc$ and $\Delta \lphinn = \lphinns-\lphinnc$
\end{itemize}
Let us point out that since the kernels are all positive functions, one has, replacing $?$ by $T$, $N$, $I$ or $F$ :
\[
\lap \phi^{?,s}_0 = ||\phi^{?,s}|| \mbox{~~and~~} \lap \phi^{?,c}_0 = ||\phi^{?,c}||,
\]
and consequently
\[
\Delta \lap \Phi^?_0 = \lap \phi^{?,s}_0 - \lap \phi^{?,c}_0 = ||\phi^{?,s}|| -||\phi^{?,c}||
\]
\end{notations}
\subsection{Stability condition - Stationarity of the price increments}
The process $P_t$ defined in Section \ref{M1} is well defined as long as the matrix $\Phi_t$ is locally integrable on $\mathds{R}^+$.
Hawkes, in his original papers \cite{Hawkes1971-1,Hawkes1971-2}, has formalized the necessary and sufficient condition for the previously introduced model  \eqref{Hawkes4d} to be stable : the matrix made of the $L^1$ norm of the elements of $\Phi$ should have eigenvalues whose modulus are strictly smaller than 1.

\noindent
This condition can be expressed in terms of conditions on the $L^1$ norm of the different kernels :
\begin{proposition}{\bf(Stability Condition)}
\label{prop:stability}
The hawkes process  $P_t$ is stable if and only if the following condition holds :
\begin{itemize}
\item[(H)] The eigenvalues of the matrix $\lPhi_0$ have a modulus strictly smaller than 1.
\end{itemize}
In that case, $P_t$ has stationary increments and the process $\lambda_t$ is strictly stationary.
Moreover (H) holds if and only if
\begin{equation}
\label{H}
c^+ < (1-a^+)(1-b^+) \mbox{~~~and~~~} a^+,~b^+ < 1,
\end{equation}
where
\begin{itemize}
\item $a^+ = \lphitts_0 + \lphittc_0$,
\item $b^+ =  \lphinns_0 + \lphinnc_0$ and
\item $c^+ =  (\lphints_0 + \lphintc_0)  (\lphitns_0 + \lphitnc_0)$.
\end{itemize}
Moreover \eqref{H} implies that
\begin{equation}
\label{H1}
\Delta \lphitt_0\Delta \lphinn_0 -1 < \Delta \lphint_0 \Delta \lphitn_0 < (1-\Delta \lphitt_0)(1-\Delta \lphinn_0),
\end{equation}
where we used Notations \ref{not:delta}
\end{proposition}
The proof is in Section \ref{proofstab}.

\noindent
\vskip .2cm
Let us point out that in the case there is no feedback of the price jumps on the trades, i.e., $\Phint = 0$ (or $c^+ = 0$), then the stability condition \eqref{H} is equivalent to
$a^+<1$ and $b^+<1$, i.e., $||\phitts|| + ||\phittc|| < 1$ and $ ||\phinns|| + ||\phinnc||<1$.

\vskip .4cm
\noindent
The mean intensity vector is given by the following Proposition.
\begin{proposition}{\bf(Mean Intensity)}
\label{prop:Lambda}
We suppose that (H) holds (i.e., \eqref{H}). Then
\begin{equation}
\label{eq:Lambda}
\Lambda = \E(\lambda_t) = (\Id -\lPhi_0)^{-1}M.
\end{equation}
This can be  written as
\begin{equation}
\label{eq:Lambdat}
\left(\begin{array}{c}
\Lambdat\\
\Lambdat
\end{array}
\right) = \mu (\Id + \lDelta_0)  (\Id-\lPhinn_0) v
\end{equation}
and
\begin{equation}
\label{eq:Lambdan}
\left(\begin{array}{c}
\Lambdan\\
\Lambdan
\end{array}
\right)
 = \mu (\Id + \lDelta_0) \lPhitn_0 v
\end{equation}
where $v = \left(
\begin{array}{c}
1 \\
 1
\end{array}\right)$ and
where $\myDelta_t$ is defined by its Laplace transform
\begin{equation}
\label{lDelta}
\lDelta_z = ((\Id-\lPhitt_z)(\Id-\lPhinn_z)-\lPhint_z\lPhitn_z)^{-1} -\Id.
\end{equation}
\end{proposition}
{\em Proof} : The proof is basically an adaptation of a proof previously presented in \cite{BDHM2011}.
Let the martingale $dZ_t$ be defined as
\[
dZ_t = dP_t -\lambda_t dt.
\]
Using  \eqref{Hawkes4d}, we get
\begin{equation}
\lambda_t = \Mu+ \Phi \star dP_t  = \Mu  + \Phi \star dZ_t+ \Phi \star \lambda_t dt.
\end{equation}
Thus
\begin{equation}
(\delta\Id-\Phi) \star \lambda_t = \Mu +\Phi \star dZ_t.
\end{equation}
Consequently,
\begin{equation}
\lambda_t = (\delta_t\Id+\Psi) \star \Mu+ (\delta_t\Id+\Psi) \star \Phi \star dZ_t,
\end{equation}
where $\Psi$ is defined by
\[
\lPsi_z =  \lPhi_z(\Id-\lPhi_z)^{-1}.
\]
Taking the expectation, we get \eqref{eq:Lambda}.
Moreover, we have
\begin{equation}
\Id-\lPhi = \left(\begin{array}{cc}
\Id-\lPhitt & -\lPhint \\
-\lPhitn  & \Id-\lPhinn
 \end{array} \right)
\end{equation}
Using Remark 1 a the end of Section \ref{M1}, one can easily check that
\begin{equation}
\label{lPhiinverse}
(\Id-\lPhi)^{-1} = (\Id + \lDelta)
\left(\begin{array}{cc}
\Id-\lPhinn & \lPhint \\
\lPhitn&\Id-\lPhitt
 \end{array} \right)
\end{equation}
where $\myDelta_t$ is defined by \eqref{lDelta}. The Equations \eqref{eq:Lambdat} and \eqref{eq:Lambdan} are direct consequences of this last equation combined with \eqref{eq:Lambda}.

\vskip .3cm
\noindent
{\bf In the following we will always consider that (H) holds, i.e., that \eqref{H} holds.}

\subsection{Numerical simulations}
\begin{figure}[h]
\centering
\includegraphics[height=8cm]{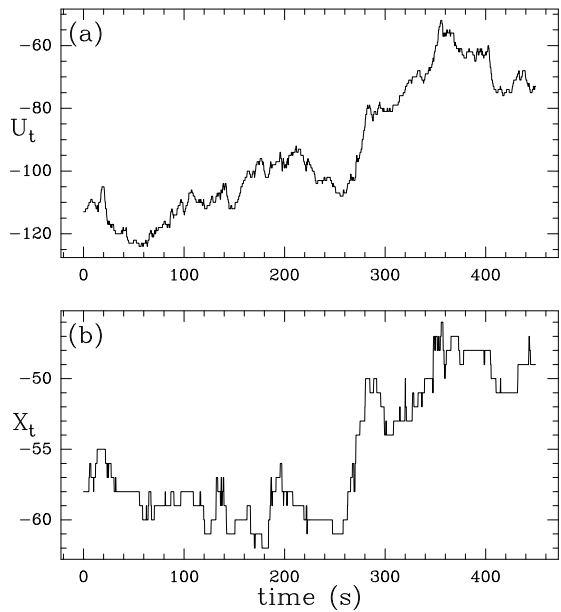}

\caption{Example of sample paths for the cumulated trade \eqref{uprocess} (a) and price \eqref{xprocess} (b) processes of the model with exponential kernels. We chose
$\phi^{T,c} = \phi^{N,s} = \phi^{I,c} = \phi^{R,s} = 0$, $\phi^{T,s} = 0.03 \; e^{-5 t/100}$, $\phi^{N,c} = 0.05 \; e^{-t/10}$, $\phi^{I,s} = 25 \; e^{- 100 t}$ and $\phi^{R,c} = 0.1 \; e^{-t/2}$.
}
\label{fig:pathExpZoom}
\end{figure}

\begin{figure}[h]
\centering
\includegraphics[height=8cm]{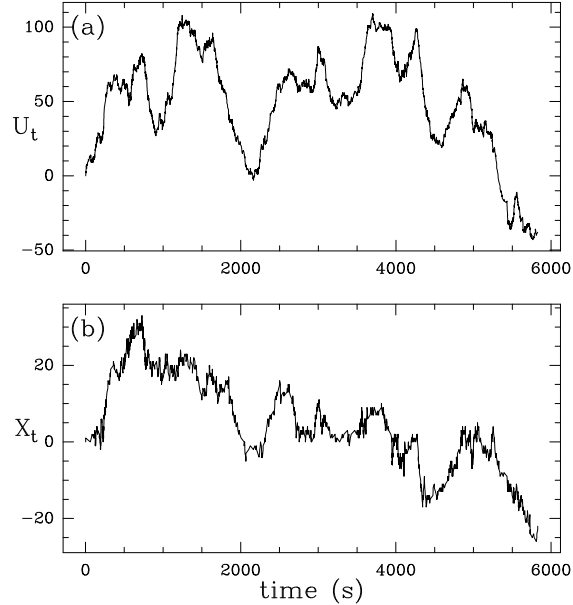}

\caption{Example of sample paths for the cumulated trade \eqref{uprocess} (a) and price \eqref{xprocess} (b) processes of the same model as in Fig. \ref{fig:pathExpZoom} over a wider range of time. One does not see the discrete nature of time
variations anymore. $U_t$ and $X_t$ appear as correlated processes.}
\label{fig:pathExp}
\end{figure}

In order to perform numerical simulations of Hawkes models, various methods
have been proposed. We chose to use a thinning algorithm (as proposed, e.g., in  \cite{Ogata1981}) that consists
in generating on $[0,t_{\max}]$ an homogeneous Poisson process with an intensity
$M > \sup_{t\in[0,t_{\max}]} (\lambda^{T\pm}_t,\lambda^{N^\pm}_t)$. A thinning procedure is then applied, each jump being accepted or rejected according to the actual value of $\lambda^{T^\pm}_t$ or $\lambda^{N\pm}_t$.
In order to illustrate the 4-dimensional process we chose to display only the price path
\begin{equation}
\label{xprocess}
X_t = N^+_t-N^-_t
\end{equation}
and the {\em cumulated trade process} path as defined by
\begin{equation}
\label{uprocess}
U_t = T^+_t-T^-_t.
\end{equation}
In Figure \ref{fig:pathExpZoom}, we show an example of sample paths of both $X_t$ and $U_t$ on a few minutes time interval.
All the involved kernels
are exponentials. Some microstructure stylized facts of the price can be clearly identified directly on the plot : price moves arrive at random times, price moves on a discrete grid and is strongly mean reverting (see beginning of next section).
In the large time limit,  one can show that
these processes converge to correlated Brownian motions (see \cite{BDHM2011} or Section \ref{diffusive}). This is illustrated in Fig. \ref{fig:pathExp} where
the paths are represented over a wider time window (almost 2 hours).
As discussed in Section \ref{empirical:correlation}, since we choose
$\phi^{T,c} = 0$ and $\phi^{N,s} = 0$, the small time increments of $U_t$ remain correlated while the price
increments correlations almost vanish. This can be observed in Fig. \ref{fig:pathExp} where the path of $U_t$ appears to be smoother
than the path of $N_t$.

In the next section we study the microstructure properties of  both the price and the market order flow.

\section{Microstructure of price and market order flow}
\label{covmatrix}
The model we introduced above can be seen as a generalization of the price microstructure model previously introduced in \cite{BDHM2010}.
 Indeed, the 2-dimensional model defined in \cite{BDHM2010}
can somewhat be seen as the "projection" on the price components $N_t$
of the model defined in Section \ref{M1}.
Thus, it easy to show that it will account for all the price microstructure stylized facts already accounted for by the model in \cite{BDHM2010}.
This  includes the characteristic decreasing shape of the {\em mean signature plot} $E((X_{\Delta t}-X_0)^2)/\Delta t$ (which explains why estimating the diffusing variance using high frequency quadratic variations leads to a systematic positive bias). As explained in \cite{BDHM2010}, this effect is  due to the high frequency strong mean reversion observed in real prices and can easily be modeled by choosing kernels such that
$||\phi^{N,c}|| \gg ||\phi^{N,s}||$.
We refer the reader to \cite{BDHM2010} for all the discussions concerning the signature plot, the problem of variance estimation using high frequency data and the link with mean reversion of price time-series.

Thus, in this Section, we shall mainly focus one price correlations and market order correlations and how they relate one to the other.
\subsection{Price and Trade covariance function}
Following \cite{BDHM2011}, we define the covariance matrix of the normalized process at scale $h$ and lag $\tau$ by
\begin{equation}\label{eq:SampledACV}
v^{(h)}_{\lag} =  h^{-1} \mbox{Cov} \left(P_{t+h+\lag}-P_{t+\lag},P_{t+h}-P_t\right) \; ,
\end{equation}
where we normalized by $h$ in order to avoid a trivial scale dependence. Let us note that, since the increments of $P_t$ are stationary (we suppose that \eqref{H} is satisfied), the previous definition does not depend on $t$. Thus, it can be rewritten as
\begin{equation}\label{eq:SampledACV1}
v^{(h)}_{\lag} =  \frac{1}{h}E\left((\int_{\lag}^{\lag+h}d{P_s}^\dagger-\Lambda h)(\int_0^{h}dP_s-\Lambda h)\right).
\end{equation}
In  \cite{BaMuDaEPJB,BDHM2011}, it is proven that the
the Laplace transform of $v^{(h)}_{\lag} $ can be expressed as a function of the Laplace transform of $g^{(h)}_{\lag}$ (where
$g^{(h)}_t=(1-\frac{|t|}{h})^+$) and of $\Phi_{\lag}$ :
\begin{equation}
\label{eq:SignaturePlotExpressionFourier}
    \widehat v^{(h)}_z =\widehat g^{(h)}_z(\Id - \widehat \Phi_z)^{-1}\Sigma(\Id - \widehat \Phi_z^\dagger)^{-1}
\end{equation}
where $\Sigma$ is the diagonal matrix
\begin{equation}
\label{sigma}
\Sigma =  \left(\begin{array}{cc}
\Lambdat \Id & 0 \\
0 & \Lambdan \Id
 \end{array} \right).
\end{equation}
From this result, one can easily deduce an analytical formula for the price auto-covariance function.

\begin{proposition}{\bf (Price auto-covariance)}
Let $C^{N,(h)}_\tau$ be the normalized auto-covariance of the price increment :
\begin{equation}
\label{covn}
C^{N,(h)}_\tau = \frac 1 h \E((X_{t+h}-X_{t})(X_{t+\tau+h}-X_{t+\tau})),
 \end{equation}
then
\begin{equation}
\label{lcovn}
 \lap{C}^{N,(h)}_z =\frac {2 \lap g^{(h)}_z(\Lambdat |\Delta\lphitn|^2+\Lambdan |1-\Delta\lphitt|^2)}
 {\left|(1-\Delta\lphitt)(1-\Delta\lphinn)-\Delta\lphitn\Delta\lphint\right|^2}
\end{equation}
\end{proposition}
Before proving this Proposition, we first need the following Lemma, which is a direct consequence of \eqref{lDelta} and \eqref{lPhiinverse}.
\begin{lem}
\label{lem1}
\begin{equation}
\label{cov}
(\Id - \widehat \Phi_z)^{-1}\Sigma(\Id - \widehat \Phi_z^\dagger)^{-1}  =  (\Id +\lDelta_z)(\Id +\lDelta^*_z) \left(\begin{array}{cc}
\lCtt_z & \lCnt_z \\
\lCtn_z & \lCnn_z \\
 \end{array} \right)
\end{equation}
where
\[
\lCtt_z = \Lambdat(\Id-\lPhinn)(\Id-\lPhinn)^*+\Lambdan\lPhint\lPhintstar,
\]
\[
\lCnt_z = \Lambdat \lPhitn(\Id-\lPhinn)^*+\Lambdan\lPhintstar(\Id-\lPhitt),
\]
\[
\lCtn_z = \Lambdat(\Id-\lPhinn)\lPhitnstar + \Lambdan(\Id-\lPhitt)^*\lPhint ,
\]
\[
\lCnn_z = \Lambdat\lPhitn\lPhitnstar+\Lambdan(\Id-\lPhitt)(\Id-\lPhitt)^*.
\]
\end{lem}
{\em Proof of Proposition : }
Using the symmetries of all the kernels, and the fact that $X_t = N^+_t-N^-_t$, we get
\[
C^{N,(h)}_\tau = \frac 2 h\left( \E((N^+_{t+\tau+h}-N^+_{t+\tau})(N^+_{t+h}-N^+_{t}))  - \E((N^-_{t+\tau+h}-N^-_{t+\tau})(N^+_{t+h}-N^+_{t}))\right)
\]
Thus, if we define $d^s_t$ and $d^c_t$ such that
\[
(\Id +\lDelta_z)(\Id +\lDelta^*_z) \lCnn_z  =
\left(\begin{array}{cc}
\lap d^s_z & \lap d^c_z \\
\lap d^c_z & \lap d^s_z \\
 \end{array} \right),
 \]
 we have
 \[
 \lap{C}^{N,(h)}_z = 2 \lap g^{(h)}_z (\lap d^s_z-\lap d^c_z)
 \]
 Using Remark 1, we get that
 \[
 \lap c^s-\lap c^c = \frac {\Lambdat |\Delta \lphitn|^2+\Lambdan |1-\Delta\lphitt|^2}
 {\left|(1-\Delta\lphitt)(1-\Delta\lphinn)-\Delta\lphitn\Delta\lphint\right|^2}
 \]
 which proves \eqref{lcovn}.

\vskip .7cm
\noindent
Using  similar computations, one derives an analytical formula for the auto-covariance of the cumulated trade process \eqref{uprocess}:
\begin{proposition}{\bf (Trade auto-covariance)}
Let $C^{T,(h)}_\tau$ be the normalized covariance of the increments of the cumulated trade process $U_t$ defined by \eqref{uprocess}.
\begin{equation}
\label{covt}
C^{T,(h)}_\tau = \frac 1 h \E((U_{t+h}-U_{t})(U_{t+\tau+h}-U_{t+\tau})),
 \end{equation}
then
\begin{equation}
\label{lcovt}
 \lap{C}^{T,(h)}_z =\frac {2 \lap g^{(h)}_z(\Lambdat |1-\Delta\lphinn|^2+\Lambdan |\Delta\lphint|^2)}
 {\left|(1-\Delta\lphitt)(1-\Delta\lphinn)-\Delta\lphitn\Delta\lphint\right|^2}
\end{equation}
\end{proposition}

\subsection{Numerical simulations}
\begin{figure}[h]
\centering
\includegraphics[height=8cm]{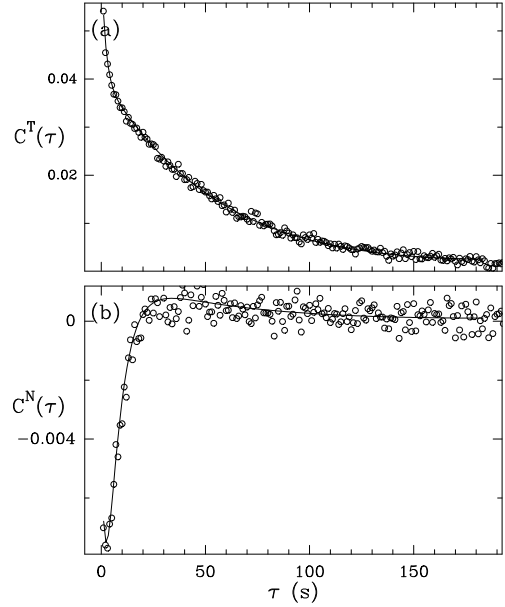}

\caption{Empirical and theoretical (From Eqs. \eqref{lcovn} and \eqref{lcovt}) autoccorelation functions for the process
defined in Fig. \ref{fig:pathExp}. (a) Autocorrelation of the increments
of $U_t$. (b) Autocorrelation of the price increments (i.e., the increments of $X_t$). In both cases, for the sake of clarity with have removed the lag $\tau = 0$.}
\label{fig:corrExp}
\end{figure}

In order to illustrate these results, in Fig. \ref{fig:corrExp}
we have plotted both theoretical (solid lines) and estimated
($\bullet$) correlation functions.
The sample we used for the estimation contains
around $300.000$ trading events and corresponds to the same kernels as the ones used for Fig. \ref{fig:pathExp}.
One clearly see that empirical estimates closely match theoretical
expressions. Moreover, one can observe
that the trade increment autocorrelation has
an amplitude that is an order of magnitude larger
than the price increment covariance. This issue will be addressed in Section \ref{empirical:correlation}.

\subsection{On the diffusive properties of the model}
\label{diffusive}
Let us point out that we know \cite{BDHM2011} that a centered $d$-dimensional Hawkes process $P_t$ diffuses at large scales ($h \rightarrow +\infty$) towards a multidimensional gaussian process :
\[
\frac 1 {\sqrt h} (P_{ht}-\E(P_{ht})) \rightarrow^{law} (\Id-\lPhi_0)^{-1} \Sigma^{1/2} W_t
\]
where $W_t$ is a standard $d$-dimensional Brownian motion and $\Sigma$ is defined by \eqref{sigma}.
This is a very general result that can be applied here to obtain the covariance matrix of the diffusive limit of our 4-dimensional Hawkes process $P_t$.
Actually, if one is just interested in the diffusive variance of $P_t$, one can easily obtain it directly from the definition of the covariance matrix
\eqref{eq:SampledACV} (noticing that $\sqrt h v^{(h)}_0$ corresponds to the variance of $\frac 1 {\sqrt h} P_{ht}$). One gets that the  diffusing variance  is $(\Id - \widehat \Phi_0)^{-1}\Sigma(\Id - \widehat \Phi_0^\dagger)^{-1}$.
It is then straightforward to obtain the diffusive variance of the price $X_t = N_t^+ - N_t^-$ :
\[
\sigma^X = \frac {\sqrt 2 \sqrt{ \Lambdat (\Delta\lphitn_0)^2+\Lambdan (1-\Delta\lphitt_0)^2}}
 {(1-\Delta\lphitt_0)(1-\Delta\lphinn_0)-\Delta\lphitn_0\Delta\lphint_0}
\]
(Let us point out that \eqref{H1} shows that the denominator is positive).

\section{Model with labeled agents}
\label{M11}
\subsection{Accounting for labeled agents}
The trade arrival process $T$ models {\em anonymous} trades sent by {\em anonymous agents}~: this corresponds to the typical trade information sent by most exchanges (i.e., the trades are {\em unlabeled}, there is no way to know whether two different trades involves the same agent or not). However, there are many situations where one has access to some {\em labeled} data corresponding to some specific {\em labeled agents}. In this case, it could be very interesting to be able to analyze the impact of these labeled trades within the same framework. It is naturally the case of brokers who have labels for all their clients (though some clients may have several brokers, consequently a given broker might not be able to identify {\em all} the trades of a given client). In general, this is the case of any financial institutions which has access to the historic of all its own trades.

In our model, we chose, to consider only the case of a
single  labeled agent  that is sending market orders at some deterministic time. The case where there are  several such agents is a straightforward generalization.
The trades arrival of this labeled agent is represented by a 2d deterministic function
\begin{equation}
A_t =  \left(\begin{array}{c} A_t^{-}  \\ A_t^{+}  \end{array} \right)
\end{equation}
representing the market orders arrival at the best ask ($A^{+}$) and at the best bid ($A^{-}$).  Again, $A^+$ (resp. $A^-$) jumps upward by 1
as soon as the agent is buying (resp. selling) (we do not take into account the associated volumes).

{\bf In all the following, we consider that the agent is active only on a finite positive time interval, i.e,
\begin{equation}
\exists T>0,~~~  Support( dA_t) \subset [0,T[
\end{equation}
}

We are ready now to reformulate the previously introduced model (see Section \ref{M1}) taking into account the labeled agent.

\subsection{The model}
\label{M2}
In its most general form the model writes
\begin{equation}
\label{LHawkes4d}
\lambda_t = \Mu + \Phi \star dP_t + \Theta \star dA_t,
\end{equation}
where $\Theta$ is a $4\times 2$ matrix.
In the same way as we did for $\Phi$, we decompose the kernel $\Theta$ using $2\times 2$ matrices :
\begin{equation}
  \Theta = \left(\begin{array}{c}
\Thetat \\ \Thetan
 \end{array} \right)
\end{equation}
\begin{itemize}
\item $\Thetat$ (influence of the labeled trades on $\lambdat$) : accounts for the herding of the anonymous trades with respect to the labeled trades.
\item $\Thetan$ (influence of the labeled trades on $\lambdan$) : accounts for the impact of a single labeled trade on the price. A priori there is no reason not to consider that the impact of a labeled trade is not the same as the impact of a anonymous trade, i.e., we will take
\begin{equation}
\Thetan = \Phitn
\end{equation}
\end{itemize}
Let us point out that the model introduced in Section \ref{M1} is a particular case (when $A_t = 0$) of this model.
In that sense, this model is a more general model.

%
%



\section{Market impact  in the model with labeled agents}
\label{sec:mi}
\subsection{Computation of the market impact profile}
Let us give an analytical expression for  the market impact profile.
\begin{proposition}{\bf (Market impact profile)}
\label{th:main}
The expectation of the intensity vector is given by
\begin{equation}
\label{eq:LambdaA}
\E(\lambda_t) = (\delta\Id +\Psi)\star (M + \Theta \star dA_t),
\end{equation}
where $\Psi$ is defined by its Laplace transform
\[
\lPsi_z =  \lPhi_z(\Id-\lPhi_z)^{-1}.
\]
The market impact profile between time 0 and time $t$ of the labeled agent is defined as
\begin{equation}
MI_t = \E(X_t) = \E(N_t^+-N_t^-).
\end{equation}
It can be written
\begin{equation}
\label{baby}
MI_t = (\delta-\Delta\xi)\star \Delta\phitn\star (A^+-A^-),
\end{equation}
where $\Delta \xi_t$ is defined as
\begin{equation}
\label{mi}
\lap{\Delta  {\xi}}_z =  1-\frac {(1-\Delta \lphitt+ \Delta \lthetat)}
 {(1-\Delta \lphitt)(1-\Delta \lphinn)-\Delta \lphitn\Delta \lphint}
\end{equation}
\end{proposition}
{\bf Proof of \eqref{eq:LambdaA}}. It is very similar to the proof of \eqref{eq:Lambda}.
Let $dZ_t$ be the martingale defined as
\[
dZ_t = dP_t -\lambda_t dt.
\]
Using  \eqref{LHawkes4d}, we get
\begin{equation}
\lambda_t = \Mu+\Theta\star dA_t + \Phi \star dP_t  = \Mu +\Theta\star dA_t  + \Phi \star \Ma dZ_t+ \Phi \star \Ma\lambda_t dt.
\end{equation}
which gives
\begin{equation}
(\delta\Id-\Phi ) \star \lambda_t = \Mu +\Theta\star dA_t + \Phi \star dZ_t,
\end{equation}
which is equivalent to
\begin{equation}
\lambda_t = (\delta\Id+\Psi) \star (\Mu+\Theta\star dA_t) + (\delta\Id+\Phi) \star \Phi_t \star dZ_t.
\end{equation}
Taking the expectation, we get \eqref{eq:LambdaA}.

\vskip .2cm
\noindent
{\bf Proof of \eqref{mi}} \\
Using \eqref{lPhiinverse},  we get
\begin{equation}
\E(\lambdan_t) = (\delta\Id+\myDelta) \star
K  \star
( \Mu+\Theta\star dA_t).
\end{equation}
where $K$ is the $2\times4$ matrix defined by
\begin{equation}
K = \left(\begin{array}{cc}
\Phitn&\delta\Id-\Phitt
 \end{array} \right)
\end{equation}
 Thus
\begin{equation}
\E(dN_t^+-dN_t^-) = u (\delta\Id+\myDelta) \star K \star
\left(\begin{array}{c}
\Thetat \star dA_t \\
\Thetan \star dA_t
 \end{array} \right)
\end{equation}
 where $u$ stands for the vector $u = (-1,1)$.
Let us recall that we chose $\Thetan = \Phitn$.
Using Remark 1,
all the $2\times2$ matrices involved in this equation are commuting since they are symmetric along both diagonals. Thus one gets
\begin{equation}
\E(dN_t^+-dN_t^-) = u (\delta\Id+\myDelta) \star (\delta  \Id - \Phitt + \Thetat) \star \Phitn \star dA_t
\end{equation}
Thus the market impact is
\begin{equation}
MI_t = u (\delta\Id+\myDelta) \star (\delta  \Id - \Phitt + \Thetat) \star \Phitn \star A_t
\end{equation}
Using Remark 1 again,
we deduce \eqref{mi}

\noindent
\vskip 1cm
Let us state a trivial corollary of this last proposition that will be of particular interest in order to study the permanent impact in the following Section:
\begin{cor}
\label{cor1}
In the case of an impulsive impact kernel (see Section \ref{impulsive}) and $\Theta^T = 0$, the market impact profile of a single buying market order sent at time $t=0$ (i.e., $dA^+_t = \delta_t$ and $dA_t^- = 0$) is given by
\begin{equation}
\label{mi11}
MI_t = 1-\int_0^t \Delta\xi_u du
\end{equation}
where $\Delta \xi_t$ is defined as
\begin{equation}
\label{mi1}
\lap{\Delta  {\xi}}_z =  1-\frac {1-\Delta \lphitt}
 {(1-\Delta \lphitt)(1-\Delta \lphinn)-I \Delta \lphint}
\end{equation}
\end{cor}

\begin{figure}[h]
\centering
\includegraphics[height=5cm]{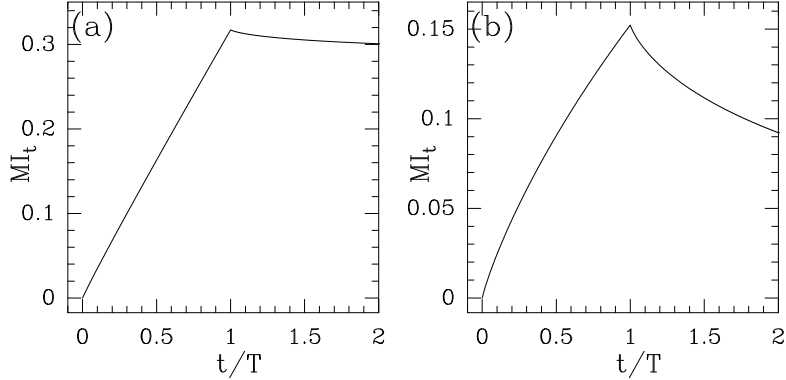}

\caption{Example of two market impact profiles according to Eq. \eqref{baby} with $\Theta^T = 0$ and
$dA_t = T^{-1} 1_{t \leq T} dt$. The kernel shapes
and the parameters have been chosen to roughly match the values observed empirically in
Section \ref{sec:empirical}. The two figures only differ by the values of $\Delta \lphitt_0$.
(a) $\Delta \lphitt_0 = 0.45$  (b) $\Delta \lphitt_0 = 0.95$.}
\label{fig:Impact}
\end{figure}

In Fig. \ref{fig:Impact} are displayed 2 examples of market impact profiles computed
according to Eq. \eqref{baby} for a meta-order consisting in buying a constant amount
of shares during a period $T$ (i.e. $dA_t = T^{-1}dt$ if $t \leq T$ and $dA_t = 0$ otherwise).
The parameters  Fig. \ref{fig:Impact}(b)
correspond to the ones
estimated from empirical data as discussed in Section \ref{sec:empirical} (notably $\Delta \lphitt_0 \simeq 0.9$).
One can see that Fig. \ref{fig:Impact}(b) reproduces fairly well the empirical shape
measured using labeled database (as, e.g., in \cite{moro2009,RaBe}): a concave shape
of the profile $MI_t$ for times $t \leq T$ (the so-called ``square-root'' law \cite{gatheral}) and a
convex relaxation towards the permanent impact when $t > T$. This shape is qualitatively explaind in Section \ref{empirical:miestimation}.

For illustration purpose, the parameters of Fig. \ref{fig:Impact}(a) are the same as the ones of Fig. \ref{fig:Impact}(b) except that they have been tweaked in order to get $\Delta \lphitt_0 < 1/2$.
The fact that the asymptotic value of the market impact is larger  is explained by point iii) of the next section.

\subsection{Permanent versus non permanent market impact}
\label{sec:permanent}
One important consequence of \eqref{mi1} is that the asymptotic market impact of a single labeled market buying order  at time $t = 0$ is of the form (assuming an impulsive market impact kernel) :
\begin{equation}
\label{pim1}
MI_{+\infty} = (1 - \Delta \lap \xi_0).
\end{equation}
Thus controlling how permanent is the market impact  essentially amounts in controlling how large $\Delta \lap \xi_0$ is
 where
\begin{equation}
\lap{\Delta  {\xi}}_0 =  1-\frac 1
 {(1-\Delta \lphinn_0)- I \Delta \lphint/(1-\Delta \lphitt_0)},
\end{equation}
thus
\begin{equation}
MI_{+\infty} = \frac 1
 {(1-\Delta \lphinn_0)-I \Delta \lphint/(1-\Delta \lphitt_0)}.
\end{equation}
Let us note that, the stability condition of the process (see Proposition \ref{prop:stability}) implies that
$|1-\Delta \lphinn_0| < 1$ and $|1-\Delta \lphitt_0| < 1$.

This last equation allows us to identify three different dynamics that can lead to a decrease of the permanent impact. They all correspond to different ways of introducing mean reversion into the price :
\begin{itemize}
\item[i)] Mean reversion due to microstructure. This case corresponds to the case $\Delta \lphinn_0$ is very negative, i.e., $||\phinnc||$ much larger than  $||\phinns||$. However this effect is limited by the fact that we know that we cannot go below $\Delta \lphinn_0>-1$ due to the stability condition.
\item[ii)] Feedback cross influence of the price moves on the trades.
This case corresponds to the case $\Delta \lphint_0$ is very negative, e.g.,
$||\phintc||$ large and  $\phints=0$. This is definitely an effect that is present in real life (see Section \ref{sec:empirical}) : as the price goes up, traders are sending more and more selling buying orders.
\item[iii)] Auto-correlation of the trades . The previous effect i) is even stronger if there is a strong correlation in the signs of the market orders (i.e., $||\Delta \lphitt_0|| \simeq 1$). Let us point out that Fig. \ref{fig:Impact} shows a clear illustration of this point : the left market impact curve has a stronger permanent market impact than the one of the right, this is due to the fact that it corresponds to a smaller  $||\Delta \lphitt_0||$.
\end{itemize}
As we will see in Section \ref{empirical:estimation}, all these effects are present in real data. And they will all play a part in reducing  the asymptotic market impact (see Section \ref{empirical:miestimation})

\subsection{Estimation from non labeled data. Response function versus Market impact function}
\label{minon}
As already pointed out, most markets do not provide labeled data. The order flows are made of anonymous orders sent by anonymous agents.
A priori, in that case, the only way of quantifying the "impact" a given market order has on the price is to estimate numerically the response function $R_t$. The response function is defined as the variation of the price from time 0 to time $t$ knowing there was a (e.g., buying) trade at time 0.
Thus it can be written
\begin{equation}
\label{response}
R_t = \E(N^+_t-N^-_t | dT^+_0 = 1), \mbox{~~for~} t>0,
\end{equation}
and $R_t = 0$ otherwise.
It can be written as
$$
R_t = \int_0^t \E(dN^+_t | dT^+_0 = 1) - \int_0^t \E(dN^-_t | dT^+_0 = 1),\mbox{~~for~} t>0,
$$
Notice that within the model \eqref{eqbouchaud} of Bouchaud et al., this response function can be
explicitly related to the ``bare impact'' function $G(n)$ and therefore used to estimate its shape \cite{bouchaud1}.
It is important to understand that this function is fundamentally different from the market impact profile of a single market order. The market impact isolates all the market orders of a single agent (e.g., a meta-order) and quantifies what the impact of these market orders are. In order to compute it, one, a priori,  needs to identify all the market orders of a particular agent.
This is not the case of  the response  function which is "polluted" by the impact of all the market orders that are in the same meta-order as the market order that is under consideration.

A very important consequence of our model (as it will be explained  in Section \ref{empirical:miestimation}) is that our model allows estimation of the market impact profile even if no labeled data are available. It will basically consist in first estimating all the kernels (see Section \ref{sec:estimation}) and then using the analytical formula \eqref{mi}.

Let us point out that one can easily obtain an analytical formula for the response function using Proposition \ref{th:cexpect} of Section \ref{annex_cond}.
We prove that if $g_t$ is the $4\times 4$ matrix defined by
$g_t = \{g^{ij}_t \}_{1\le i,j \le 4}$ with
\[
g^{ij}_t dt = \E(dP^i_t| dP^j_0 = 1) - \epsilon^{ij} \delta_t -\Lambda^i dt,
\]
then
\[
\lap g_z = (\Id-\lPhi_z)^{-1}\Sigma (\Id-\lPhi_z^\dagger)^{-1}\Sigma^{-1} -\Id.
\]
Using Lemma \ref{lem1}, if we define
\[
R^s_t = \int_0^t \E(dN^+_u | dT^+_0 = 1) -\Lambdan du \mbox{~~~and~~~} R^c_t = \int_0^t  \E(dN^-_u | dT^+_0 = 1)-\Lambdan du,
\]
and if we define $dR^s_t = r^s_t dt$,  $dR^c_t = r^c_t dt$,
then,
\[
\left(
\begin{array}{cc}
\lap r^s_z\\
\lap r^c_z
\end{array}
\right) = (\Id + \lDelta)(\Id + \lDelta^*) (\lCtn)^{*}/ \Lambdat
\]
Using Remark 1, we get
\[
\lap r_z = \lap r^s_z - \lap r^c_z =  \frac {(\Id-\Delta \lphinnstar)\Delta \lphitn + (\Lambdan/\Lambdat)(\Id-\Delta \lphitt)\Delta \lphintstar}
 {\left|(1-\Delta\lphitt)(1-\Delta\lphinn)-\Delta\lphitn\Delta\lphint\right|^2},
\]
In the case of an impulsive impact kernel, it gives
\[
R_t =  I (1 - \int_0^t \Delta \xi'_u  du),~~~\forall t > 0
\]
where $\Delta \xi'_t$ is defined by
\[
\lap {\Delta \xi'}_z =  1-\frac {(1-\Delta \lphinnstar) + (\Lambdan/\Lambdat)(1-\Delta \lphitt)\Delta \lphintstar I^{-1}}
 {\left|(1-\Delta\lphitt)(1-\Delta\lphinn)-I\Delta\lphint\right|^2}
\]

\section{Non parametric estimation of the kernel functions}
\label{sec:estimation}
\begin{figure}[h]
\centering
\includegraphics[height=9cm]{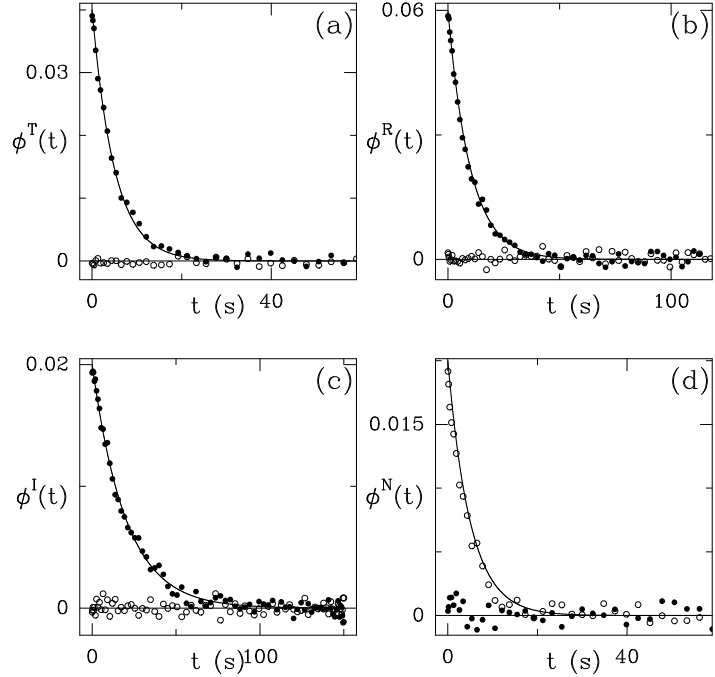}

\caption{Numerical estimation of the Hawkes kernel matrix $\phi$ by solving the Fredholm
equation $\eqref{fredholm}$ for an exponential model. Estimations are represented by symbols
$(\bullet)$ for the ``self'' kernels and $(\circ)$ for the ``cross'' kernels.
Exact exponential kernels are represented by the solid lines. The model corresponds to
$\phi^{T,c} = \phi^{N,s} = \phi^{I,c} = \phi^{R,c} = 0$,
$\phi^{T,s} = 0.04 \; e^{-t/5}$, $\phi^{N,c} = 0.02 \; e^{-t/5}$,
$\phi^{I,s} = 0.02 \; e^{-t/20}$ and $\phi^{R,s} = 0.06 \; e^{-t/10}$.}
\label{fig:estimExp}
\end{figure}
In this section we provide a new method to estimate the shape
of the kernels involved in the definition of the model\footnote{Notice that during the completion of this paper, we have been
aware of a work by M. Kirchner introducing
a non parametric estimation method very similar to the one presented in this section.
Let us point out that he has performed a comprehensive statistical study of the main
properties of this estimator \cite{Kirch012}.}.
A former non parametric estimation method has been introduced in
\cite{BaMuDaEPJB}. This method relies on the expression \eqref{eq:SignaturePlotExpressionFourier} and mainly consists in
extracting the square root of the autocorrelation matrix.
However, in order to do so univocally, one has to suppose that
the process is fully symmetric and in particular that
$T$ and $N$ have the same laws.
This assumption is clearly unrealistic, therefore the
method of \cite{BaMuDaEPJB} is not suited to estimate
the process defined in Section \ref{sec:defmodel}. For that reason we
introduce an alternative method that does not require any symmetry hypothesis. This method relies on the Proposition \ref{th:fredholm} of Annex \ref{annex_cond} (Eq. \eqref{fredholm}):
\[
g_t = \phi \star (\delta\Id + g_t),~~~~\forall t > 0.
\]
where $g_t$ is the matrix of conditional expectations defined
in Eq. \eqref{gij}. This above equation is a Fredholm equation
of 2nd kind. Since $g_t$ can be easily estimated from empirical data,
the matrix $\phi$ can be obtained as a numerical solution of the Fredholm
system. We thus implemented a classical Nystrom method that amounts to
approximate the integrals by a quadrature and solve a linear system
\cite{numrecipes}.
\begin{figure}[h]
\centering
\includegraphics[height=5cm]{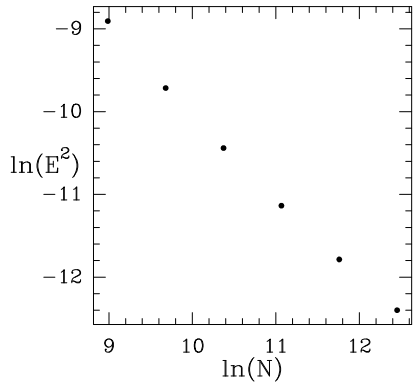}

\caption{Estimation squared error as a function of the number of events
 for the model of Fig. \ref{fig:estimExp} in log-log scale.
 The slope is $-1$.}
\label{fig:estimError}
\end{figure}
In order to illustrate the method, we have reported in Fig. \ref{fig:estimExp}
the estimated kernels in the case when all functions $\phi_t$ have
an exponential shape. The sample used has typically $3 \; 10^5$ trade events and $1.5 \; 10^5$ price change events.
One can see that for such sample length, numerical estimates
and the real kernels are close enough to determine a good fit of the latter.
Let us note that we have checked that the method is reliable
for various examples of kernels like exponential, power-laws
or constant over an interval.

In order to investigate the efficiency of the method, one can evaluate
the estimation error behavior as a function of the number of events.
This error can be defined as the supremum
of the mean square error associated with each kernel:
if $\tilde{\phi_{N_e}}$ stands for the estimated
kernel matrix for $N_e$ trading events, then one can consider
the following square error:
\[
     e^2(N_e) = E \left( \sup_{i,j} ||\phi_{ij} -\tilde{\phi}_{ij} ||^2 \right)
\]
A log-log plot of $E^2(N_e)$ as a function of $N_e$ is reported in
Fig. \ref{fig:estimError} where $E^2(N_e)$ has been estimated, for each $N_e$, using 500 trials of the model.
The measured slope is close to $-1$ in agreement with the standard behavior of the estimation error:
\[
   e(N_e) \simeq \frac{E_0}{\sqrt{N_e}} \; .
\]
A comprehensive study of the statistical properties of the
method is out of the scope of the present paper and will be addressed
in a forthcoming work.

\section{Application to real data}
\label{sec:empirical}
In this section we apply the main theoretical results we obtained in the previous sections to real data.
We consider intraday data associated with the most liquid
maturity of EuroStoxx (FSXE) and Euro-Bund (FGBL) future contracts.
The data we used are trades at best bid/ask
provided by QuantHouse Trading Solution \footnote{http://www.quanthouse.com}.
Each time series covers a period of 800 trading days going from 2009 May to 2012 September.
The typical number of trades is around 40.000 per day while the number of mid price changes is 20.000 per day.

\subsection{Kernel matrix estimation}
\label{empirical:estimation}

\begin{figure}[h]
\centering
\includegraphics[height=8cm]{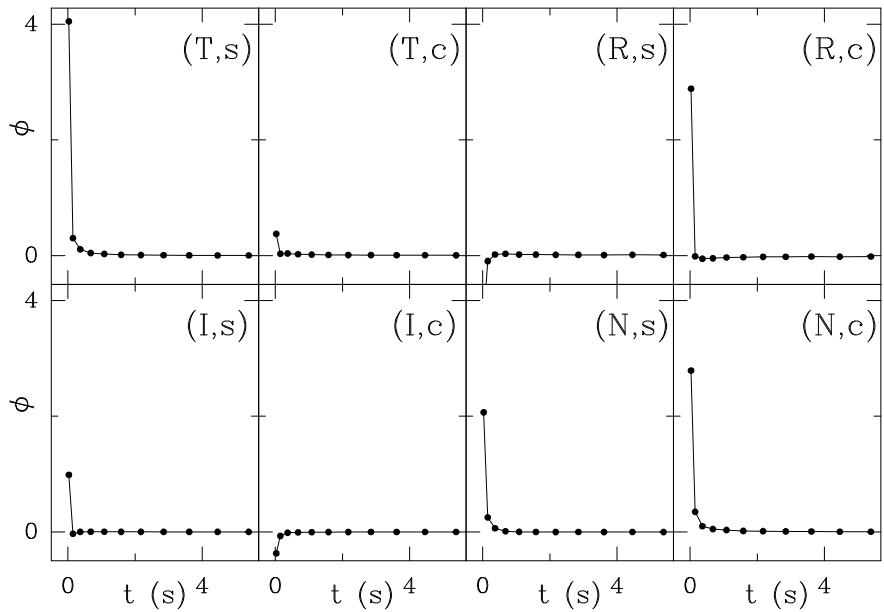}

\caption{Numerical estimation of the Hawkes kernels of EuroStoxx in the intraday slot [9 a.m., 11 a.m.]}
\label{fig:phisxeF}
\end{figure}

\begin{figure}[h]
\centering
\includegraphics[height=5cm]{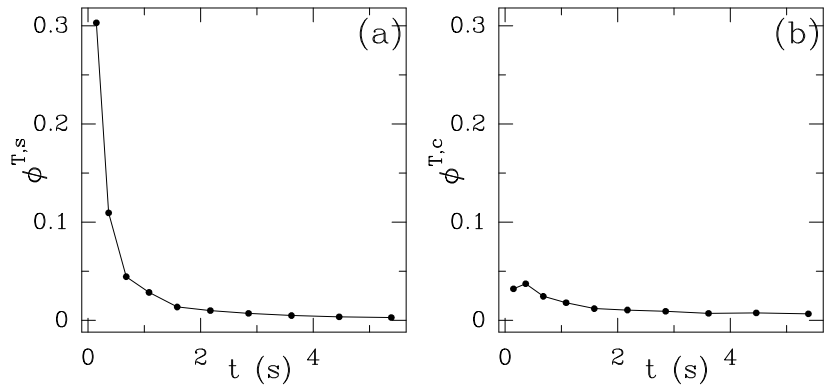}

\caption{Numerical estimation of the Hawkes kernels $\phi^{T,s}$ (a)
and $\phi^{T,c}$ (b) in the intraday slot [9 a.m., 11 a.m.] for EuroStoxx.
This is a ``zoom'' of Fig. \ref{fig:phisxeF}. One sees that $\phi^{T,s}$
slowly decreases and is very large as compared to $\phi^{T,c}$.}
\label{fig:phitsxe}
\end{figure}

\begin{figure}[h]
\centering
\includegraphics[height=5cm]{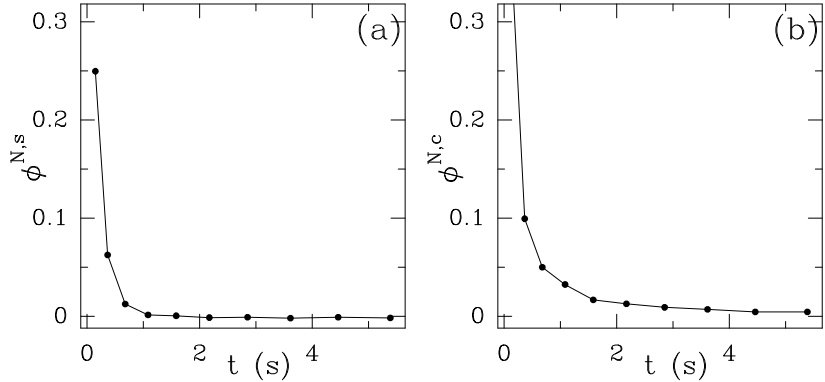}

\caption{Numerical estimation of the Hawkes kernels $\phi^{N,s}$ (a)
and $\phi^{N,c}$ (b) in the intraday slot [9 a.m., 11 a.m.] for EuroStoxx.
This is a ``zoom'' of Fig. \ref{fig:phisxeF}. One sees that $\phi^{N,c}$
is larger than $\phi^{N,s}$.}
\label{fig:phinsxe}
\end{figure}

\begin{figure}[h]
\centering
\includegraphics[height=5cm]{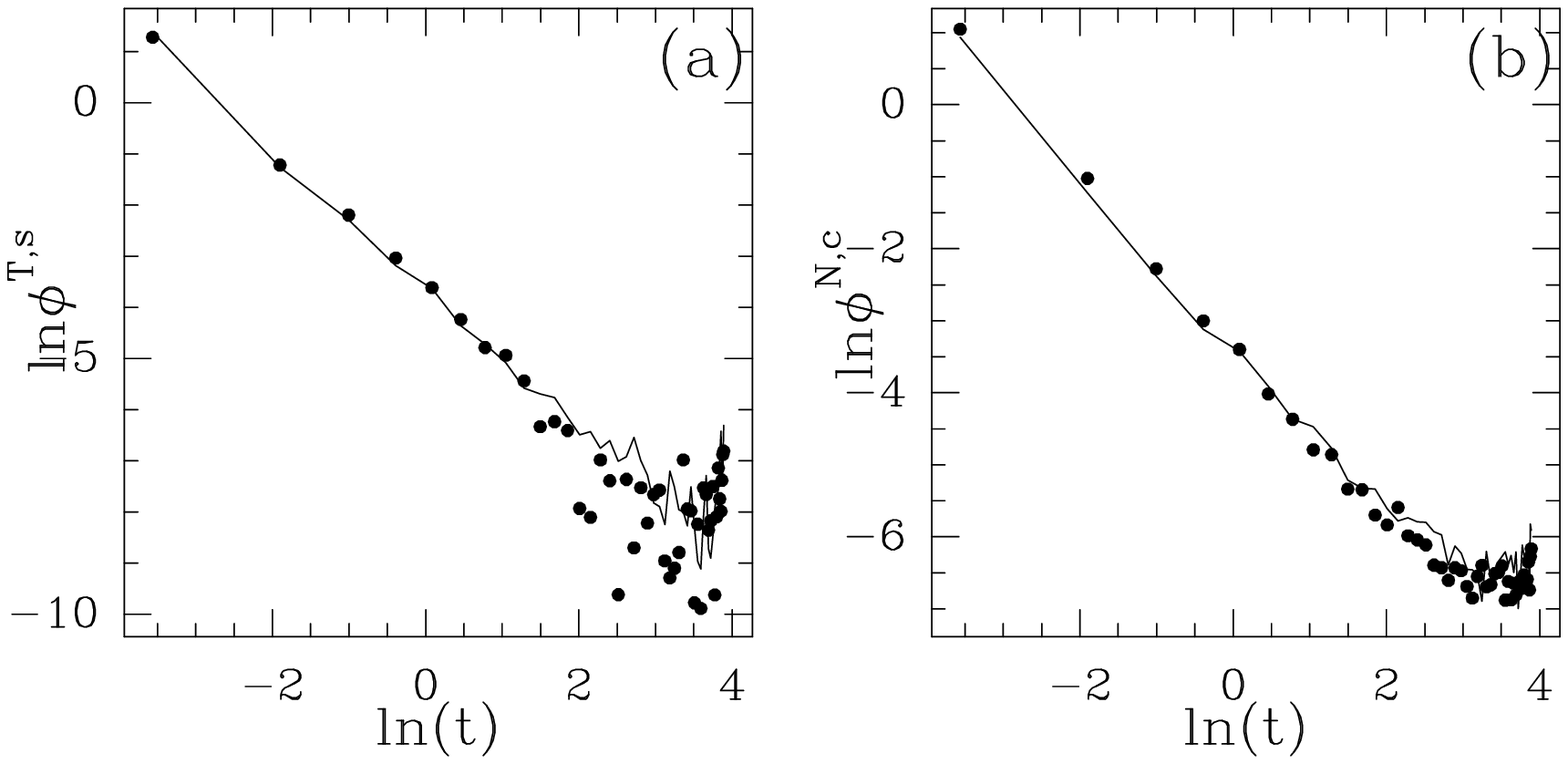}

\caption{Scaling of Hawkes kernels $\phi^{T,s}$
and $\phi^{N,c}$. The kernels are represented in double
logarithmic representation for Eurostoxx ($\bullet$) and
Euro-Bund (solid lines) estimates in the intraday slot [9 a.m., 11 a.m.]. (a) $\phi^{T,s}$ (b) $\phi^{N,c}$.
These plots suggest that the kernels have the same power-law behavior for both EuroStoxx and Euro-Bund.}
\label{fig:philoglogsxefgbl}
\end{figure}

In this section we apply the non parametric kernel estimation algorithm presented in Section \ref{sec:estimation} to our data.
Since intraday statistics are well known to be non stationary,
estimations are based on a 2 hours liquid period from 9 a.m. to 11 a.m. (GMT).
The results of the kernel estimation of EuroStoxx  are displayed in
Figs.\ref{fig:phisxeF}, \ref{fig:phitsxe} and \ref{fig:phinsxe}.

One first sees that $\phi^{T,c}$
is small as compared to $\phi^{T,s}$. This appears more
clearly on  the ``zoom'' presented in Fig. \ref{fig:phitsxe} where we have removed the first point in order to get a finer
scale. The fact that $\phi^{T,s}$ is larger than $\phi^{T,c}$ confirms the well known strong correlation observed in trade signs
(see Section \ref{empirical:correlation}).
The mid-price dynamics is well known to be mainly
mean-reverting \cite{BDHM2010}. This is confirmed by
Fig. \ref{fig:phinsxe} where a greater intensity of the kernel $\phi^{N,c}$ as compared to $\phi^{N,s}$ can be observed.
Both kernels $\phi^{T,s}$ and $\phi^{N,c}$ are found to decrease
as a power-law:
\[
   \phi_t = \alpha \left(c+t \right)^{-\beta}
\]
with an exponent $\beta \simeq 1.2$ for $\phi^{T,s}$ and
$\beta \simeq 1.1$ for $\phi^{N,c}$. The cut-off $c$
insures the finiteness of the $L^1$ norm of $\phi$ and is found
to be smaller than $10^{-2}s$  while the values of $\alpha$
are typically between $0.05$ and $0.1$.
This power-law behavior is illustrated
in Fig. \ref{fig:philoglogsxefgbl} where we have reported
in log-log scale the estimates $\phi^{T,s}$ and $\phi^{N,c}$
for both EuroStoxx ($\bullet$ ) and Euro-Bund (solid lines).
Let us notice that power-law behavior of Hawkes kernels
have already been observed in \cite{BaMuDaEPJB} using
a simple 2D Hawkes model of price jumps. This important property suggests the
existence of some scale invariance properties underlying
the order book dynamics.
Let us point out that Fig. \ref{fig:philoglogsxefgbl}
also suggests that the kernels are the same for
EuroStoxx and Euro-Bund.

As far as the impact is concerned, one can see in Fig. \ref{fig:phisxeF}
that only $\phi^{I,s}$ is significant and well modeled by an
impulsive kernel (see Section \ref{impulsive}). This confirms the fact that
a buy (resp. sell) market order increases the probability of an
upward (resp. downward) movement of the mid-price but only within
a very small time interval after the trade.
The feed-back kernel that
accounts for the influence of a price move on the trading
intensity is also found to be well localized but only
the cross term is non negligible. It seems that an upward
(resp. downward) move of the mid price triggers
an higher trading activity on the bid side (resp. on the ask side).


\subsection{Introducing a model with intraday seasonalities}
\label{empirical:seasestimation}
\begin{figure}[h]
\centering
\includegraphics[height=9cm]{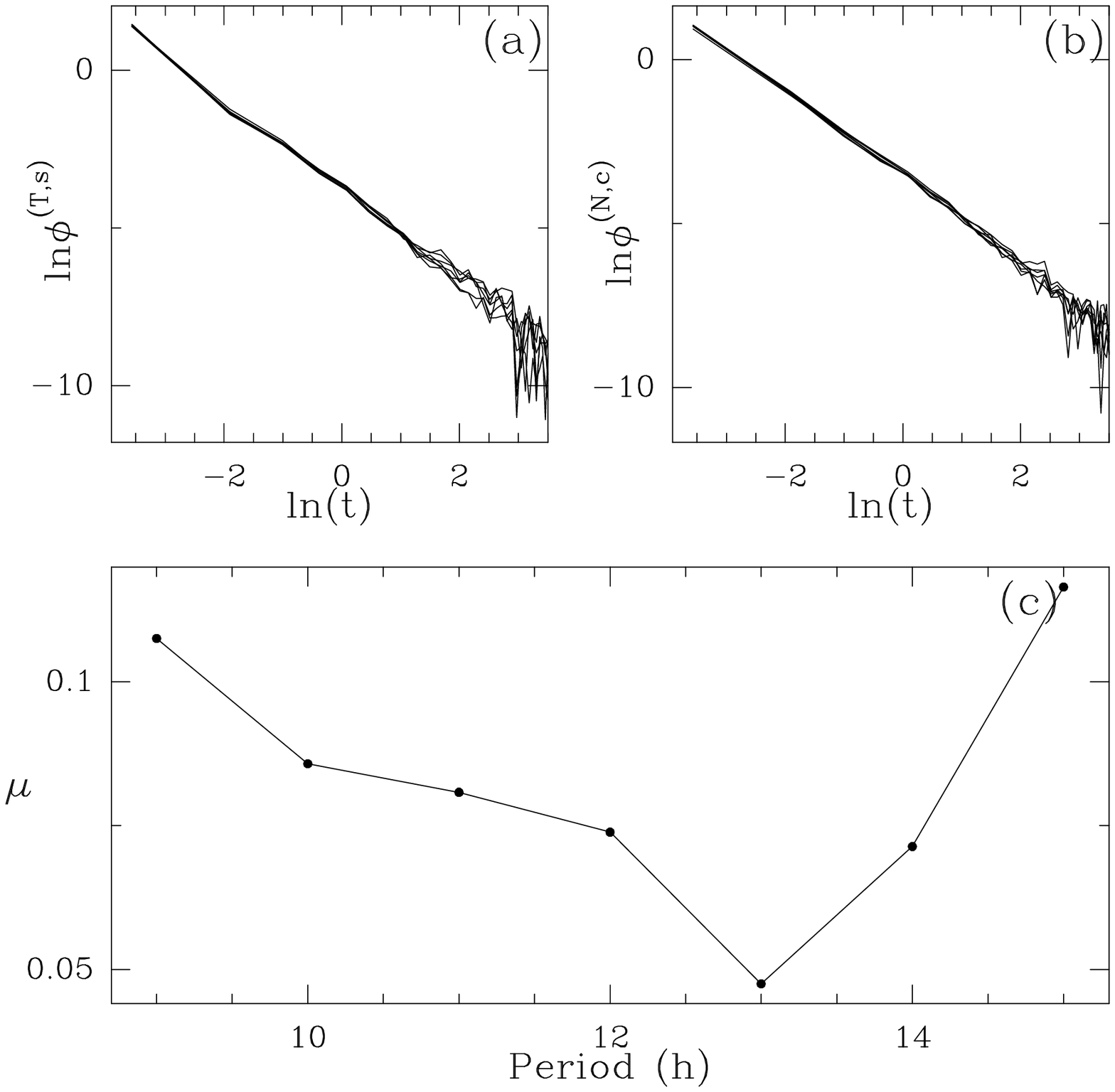}

\caption{Numerical estimation of the Hawkes kernel matrix $\phi$
for various 2 hours slices during the day for EuroStoxx.
In (a) and (b) 7 estimated kernels are represented in doubly
logarithmic scale corresponding to the 7 slots: [8 a.m., 10 a.m.], [9 a.m., 11 a.m.], [11 a.m., 1 p.m],
[12 a.m., 2 p.m.], [1 p.m., 3 p.m], [2 p.m., 4 p.m] and [3 p.m., 5 p.m.]. In (a) $\phi^{T,s}$ and in (b) $\phi^{N,c}$. One sees
that the kernel shapes are remarkably stable power-laws across intraday time periods. In (c) we reported
the estimation of the constant rate $\mu$ as a function of the intraday time. One recognizes
the classical U-shaped behavior.}
\label{fig:stability}
\end{figure}
In order to check the stationarity of the kernels, we performed various estimations
on different intraday slices of 2 hours. The results for $\phi^{T,s}$ and $\phi^{N,c}$
are reported in Fig. \ref{fig:stability}(a) and Fig. \ref{fig:stability}(b)
in the case of EuroStoxx. One can see that
the shape of the kernels does not depend on the intraday market activity and are remarkably
stable throughout the day. An estimation of the norms of all kernels allows one
to estimate the rate $\mu$ through \eqref{eq:Lambdat} or \eqref{eq:Lambdan}. In Fig. \ref{fig:stability}(c),
we see that this parameter follows the classical U-shaped intraday curve.
It thus appears that, in the model,
the exogenous intensity of trades fully accounts for the intraday modulation of market activity.

It is therefore natural to consider the following model with seasonal variations that
generalizes, in a particularly simple way, the definition \eqref{Hawkes4d}:
\begin{equation}
\label{Hawkes4dSeas}
\lambda_t = \Mu_t + \Phi \star dP_t,
\end{equation}
with
\begin{equation}
\Mu_t =  \left(\begin{array}{c}
\mu_t \\
\mu_t \\
0 \\
0
 \end{array} \right).
\end{equation}
where $\mu_t$ is a U-Shaped 1-day periodic function.
Let us point out that this is particularly elegant and simple way of accounting for an a priori pretty complex phenomenon.

\subsection{Trades correlations, Price correlations and  Efficiency}
\label{empirical:correlation}
Few years ago,  Bouchaud et al. \cite{bouchaud1} and Lillo and Farmer \cite{FarmerLillo04}
independently brought evidence that the market order flow is a long memory process.
By studying the empirical correlation function (in trading time) of the signs of trades they have shown, for different markets and assets that:
\begin{equation}
\label{covtrades}
   C(n) \sim n^{-\gamma}
\end{equation}
where $n$ is the lag in number of trades (i.e., using trading time).
This scaling law remains valid over 2 or 3 decades and the exponent $\gamma$ is in the interval $[0.4,0.7]$.
The aim of impact price models as described in the introductory section (Eq. \eqref{eqbouchaud})
was to solve this "long-memory" puzzle.
Indeed, since trades impact prices, if trades are long-range correlated, in order to maintain efficiency,
the price have to respond to market order through
a long-memory kernel. This is precisely the meaning of the kernel $G(j)$
in \eqref{eqbouchaud}. Bouchaud et al. \cite{bouchaud1} have shown that, provided
the specific shape of $G(j)$ is adjusted as respect
to the behavior \eqref{covtrades}, trade correlations impact on prices can be canceled
(see also \cite{BFL09} for an interpretation of this price model in terms of prediction error or ``surprise'').

In this section, we explain how this issue can be addressed within our model. It is important to point out that, as we will see,
the apparent long-range correlation of the trade signs and the price efficiency will be a consequence of 3 empirical findings (cf Section \ref{empirical:estimation}):
\begin{itemize}
\item $\Delta\phitt_t$ is power-law
\item $\Delta\lphitt_0$ is close to 1, i.e., it almost saturates the stability condition $\Delta\lphitt_0 < 1$
\item $\Delta\lphinn_0 < 0$, i.e., the price at high frequency is mainly mean-reverting.
\end{itemize}

\vskip .5cm
\paragraph{{\bf About the long-range memory of the trade sign process}}
The exact expression of the
autocorrelations of the increments of $U_t=T^+_t-T^-_t$ and $X_t=N^+_t-N^-_t$ as a
function of the lag $\tau$ can be hardly deduced from \eqref{lcovn}
and \eqref{lcovt}.
In order to discuss the shape of theses correlation functions
one can however use qualitative arguments based on classical
Tauberian theorems \cite{Feller}. In what follows the argument
$z$ of Laplace transforms is assumed to be real and positive ($z>0$).

Let us first remark that, within
the range of parameters observed in empirical data (and notably the relationship $\Lambda^T \simeq 2 \Lambda^N$),
one can show that the terms involving $\Delta \lphint$ both at the numerator and at the denominator in \eqref{lcovt} are subdominant.
In this case, this equation,
reads (we neglect the $g^{(h)}$ factor and drop the $(h)$ upper-script everywhere):
\begin{equation}
\label{lcovtsimple}
\lap{C}^{T}_z \simeq \frac {2 \Lambdat}
 {\left|(1-\Delta\lphitt_z)\right|^2}
\end{equation}
Let us note that, according to this expression, the stability condition
$\Delta \lphitt_0 < 1$ implies that
$\lap{C}^{T}_0 < \infty$, i.e.,
the correlation function of the trades $C^{T}_\tau$ is integrable and that, strictly speaking, there is no long-range memory
in supply and demand. Let us show however that, under the conditions we observe
empirically, $C^{T}_\tau$  can reproduce, on a pretty large (though finite) range of $\tau$, a slow-decay with an exponent smaller than 1,
leading various numerical estimations to conclude long-range memory.

As observed in previous section, $\Delta \phi^{T}$ is close to a power-law ($t > 0$):
\begin{equation}
\label{eq:deltaphipowerlaw}
  \Delta \phi^{T}_t = \alpha (c+t)^{-\beta}
\end{equation}
were $\beta = 1+\nu$ is the scaling exponent (empirically we found $\nu \simeq 0.2$),
$c$ is a small scale cut-off (empirically $c \leq 10^{-2}s$),
and $\alpha$ is a factor such that the norm $\Delta \lphitt_0 \leq 1$.
On has obviously:
\[
  \alpha = \frac{(\beta-1) \Delta \lphitt_0}{c^{1-\beta}}
\]
If one computes the Laplace transform (with $z \geq 0$) of expression \eqref{eq:deltaphipowerlaw},
it is easy to show that, in the limit of small $z$ (in practice that means $z < c^{-1}$)
\[
  \Delta \lphitt_z \simeq \Delta \lphitt_0 \left(1-\Gamma(1-\nu) (cz)^{\nu} \right)
\]
Thus, according
to \eqref{lcovtsimple}, one has
 \begin{equation}
 \label{eq:deltaphipowerlaw1}
 \lap{C}^{T}_z \simeq \frac {2 \Lambdat}
 {\left|(1-\Delta\lphitt_0) + \Delta\lphitt_0\Gamma(1-\nu) (cz)^\nu\right|^2}
\end{equation}
and consequently
$
 \lap{C}^{T}_z \sim C - C' z^{\nu}
$
and therefore, thanks to Tauberian Theorems (limit of small $z$ corresponds to limit of large time) :
\[
  C^{T}_\tau \operatornamewithlimits{\sim}_{\tau \rightarrow \infty} \tau^{-\beta}
\]
However, if we suppose that not only $z$ is small but that $\Delta \lphitt_0$ is close enough to 1 such that
\begin{equation}
\label{ineqr}
1- \Delta \lphitt_0 \leq  \Delta \lphitt_0  \Gamma(1-\nu) (c z)^{\nu}
\end{equation}
then \eqref{eq:deltaphipowerlaw1} becomes
\[
  \lap{C}^{T}_z \sim z^{-2 \nu}
\]
or equivalently
\begin{equation}
\label{interms}
  C^{T}_\tau \sim \tau^{2 \nu -1}
\end{equation}
Let us point out that the inequality \eqref{ineqr} holds, as long as:
\begin{equation}
  z >  c^{-1} \left( \frac{1-\Delta \lphitt_0}{\Delta \lphitt_0 \Gamma(1-\nu)} \right)^{\frac{1}{\nu}} \simeq  c^{-1} 10^{-5}
\end{equation}
using the estimates $\nu \simeq 0.2$ and $\Delta \lphitt_0 \simeq 0.9$.
Since $z < c^{-1}$, this means
that it can take 5 decades to see the ``short range'' nature of the
trade correlation function. Considering that $c \simeq 10^{-2} s$,
for an inter event mean time of $1 s$, the scaling \eqref{interms} can
extend over 3 decades.
Since $1-2\nu \simeq 0.6$, this exponent is precisely in the range of
values reported empirically in \cite{bouchaud1} and \cite{FarmerLillo04}.

\vskip .5cm
\paragraph{{\bf About price efficiency}}
From an empirical point of view, in agreement with market efficiency hypothesis,
it is well known that price variations
are almost uncorrelated after few seconds.
Let us show, that, under the conditions observed in empirical
data, $C^{N}_\tau$ decreases very fast around $\tau = 0$.
We provide the same kind of pedestrian arguments as in previous discussion:
we neglect the influence of $\Delta \phi^F$
and suppose that we are in the ``impulsive'' case
of the impact kernel, i.e., $\Delta \phi^{I}_t = I \delta_t$.
Let us remark that, since $\Lambda^T/\Lambda^N$ is bounded (empirically $\Lambda^T / \Lambda^N \simeq 2$), if $I$ is small enough
(empirically $I < 10^{-1}$), the same arguments invoked for $C^{T}$
shows that in the intermediate range:
\[
            c^{-1} 10^{-5}   \leq z \leq c^{-1}
\]
one has $\Lambda^T |\Delta \lphitn_z|^2 \leq \Lambda^N |1-\Delta \lphitt|^2$.
It results that \eqref{lcovn} reduces, in this range, to:
\[
\lap{C}^{N}_z \simeq \frac{2 \Lambda^N}
 {\left|(1-\Delta \lphinn_z)\right|^2}
\]
In order to study the behavior of this function, we use again
a power-law expression for $\Delta \phi^N_t$:
\begin{equation}
\label{eq:deltaphinpowerlaw}
  \Delta \phi^{N}_t = \alpha' (c+t)^{-\beta'}
\end{equation}
where empirically $\alpha' < 0$ (and thus $\Delta \lphinn_0 < 0$)
and $\beta' = 1+\nu'$ with $\nu' \simeq 0.1$.
It follows that
\[
 \lap{C}^{N}_z \simeq C
\]
in the range
\[
   z \ll c^{-1} \left(\frac{1-\Delta \lphinn_0}{\Gamma(1-\nu') |\Delta \lphinn_0|}\right)^{1/\nu'}
\]
Since $\Delta \lphinn_0 < 0$, $|\Delta \lphinn_0| < 1$, $\Gamma(1-\nu') \simeq 1$, $1/\nu' \simeq 10$,
we have $1 \ll \left(\frac{1-\Delta\lphinn_0}{\Gamma(1-\nu') |\Delta\lphinn_0|}\right)^{1/\nu'}$
and the previous inequality always holds.
This shows that the price increments correlation functions decreases very fastly around zero without any ``fine tuning''
of the kernel shapes.

\begin{figure}[h]
\centering
\includegraphics[height=5cm]{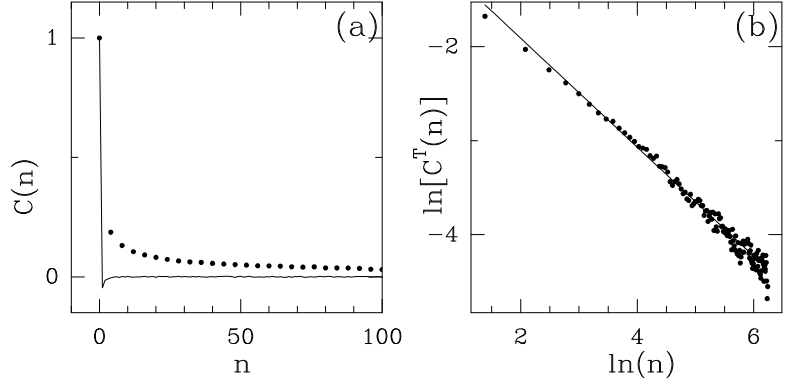}

\caption{Empirical correlation functions of the increments $U_t$ and $X_t$ in trading
time. The process is a numerical simulation of a Hawkes process where the kernels
have been chosen to fit the empirical observations found in \ref{empirical:estimation}. The correlation functions
where estimated on a single realization containing around 40.000 market orders.
(a) $C^T_n/C^T_0$ ($\bullet$) and $C^N_n/C^N_0$ (solid line) as functions of the lag $n$ (expressed in trading time).
(b) $C^T_n$ in double logarithmic reprensentation ($\bullet$). The solid line
represents the power-law fit with expression \eqref{interms} where $2 \nu-1 = -0.6$.}
\label{fig:covTNmodel}
\end{figure}

\paragraph{{\bf Numerical illustrations}}
All these considerations are illustrated in Fig. \ref{fig:covTNmodel}, where we have simulated
a Hawkes process with the parameters close to the ones we found empirically.
Let us notice that, in order to mimic the experiments performed in \cite{bouchaud1} and \cite{FarmerLillo04},  the correlation functions are computed in trading time, i.e., a discrete time which is incremented by 1 at each jump of $U_t$. We checked that setting a lag $n$ in trading time roughly amounts to
consider a lag $\tau$ in physical time such that: $\tau = \frac{n}{\Lambda^T}$.
In that respect, the shapes of the correlation functions in trading time and in physical
time are very close to each other, up to a scaling factor.
In Fig. \ref{fig:covTNmodel}(a) are displayed $C^T_n/C^T_0$ ($\bullet$) and $C^N_n/C^N_0$ (solid
line) as functions of the lag $n$ (in trading time). On clearly sees that $C^T_n$ is slowly decaying
while $C^N_n$ almost vanishes after few lags. In Fig. \ref{fig:covTNmodel}(b) $C^T_n$ is represented
in log-log. As expected, it behaves as power-law with the exponent $2\nu-1$ ($-0.6$ in the example we
choose) represented by the solid line.

\subsection{Market impact profile estimation from anonymous data}
\label{empirical:miestimation}
As discussed in Section \ref{sec:mi}, once one has
estimated all the model parameters, it is possible to compute
the shape of the market impact of some particular (meta-) order
(Proposition \ref{th:main} with $\Theta^T = 0$). If one uses the parameters reported
previously, one gets a shape of the market impact profile associated with
the meta-order $dA_t = T^{-1}1_{t\leq T}dt$ similar to the one
displayed in Fig. \ref{fig:Impact}(b).
One can use similar qualitative arguments than in previous section to explain
the shape observed in Fig. \ref{fig:Impact}(b).
Indeed, according to \eqref{baby} and using the same notations, parameter values (notably the fact that
$\Delta \lphitt_0$ is close to 1) and approximations
as discussed previously, one can show that in a wide intermediate range of laplace parameters $z$, we have:
\begin{equation}
  {\lap MI}_z \simeq z^{\nu} {\lap A}_z.
\end{equation}
If $dA_t = T^{-1} 1_{t \leq T}dt$, then ${\lap A}_z = \frac{1-e^{-zT}}{T z^2}$ and therefore
\begin{eqnarray*}
   {\lap MI}_z & \sim & z^{\nu-1} \; \; \mbox{if} \; \; z \ll T^{-1} \\
   {\lap MI}_z & \sim & z^{\nu-2} \; \; \mbox{if} \; \; z \gg T^{-1} .
\end{eqnarray*}
It follows that the market impact profile behavior reads:
\begin{eqnarray*}
   MI_t & \sim & t^{-\nu} \; \; \mbox{if} \; \; t \gg T  \\
   MI_t & \sim & t^{1-\nu} \; \; \mbox{if} \; \; t \ll T.
\end{eqnarray*}
that corresponds to the behavior observed in Fig. \ref{fig:Impact}(b).
Notice that a strict ``square-root'' would correspond to $\nu = 1/2$ while
it seems we rather observe $\nu \simeq 0.2$ empirically.

In general, the determination
of the market impact profile is a hard task that requires to possess
agent labeled (e.g. broker) data. The model presented in this paper
allows one to recover a market profile using anonymous order flow data as explained in Section \ref{minon}.

\section{Conclusion and prospects}
\label{conclusion}
From our knowledge, the model we developed in this paper is the first model
that accounts for both stylized facts of market prices microstructure (including
random time arrival of price moves,  discrete price grid, high frequency mean reversion, correlation
functions behavior at various time scales) and the stylized facts of
market impact (mainly the concave/relaxation characteristic shape of the market impact of a meta-order).
Analytical closed formula can be obtained for most of these stylized facts.
Not only it allows us to reveal (through the estimations of the different kernels)
the dynamics involved between trade arrivals and price moves, but it also allows us to estimate the entire market impact profile from anonymous market data.

As far as trade and price dynamics are concerned, we have provided evidence
of a power-law behavior of the kernels $\phi^{T}$ and $\phi^{N}$ and that the model is close to instability (i.e. $\Delta \lphi^T_0$ is smaller but close to 1).
 This suggests the existence
of some self-similarity properties in the order-book dynamics and sharply contrasts
with the usual exponential kernels used in former parametric Hawkes modeling in Finance.
The cross-kernels associated with impact of trade on prices ($\phi^I$) and
feed-back ($\phi^F$) are well localized in time (i.e. of ``impulsive nature''). Thus,
upward (resp. downward) price moves are mainly impacted by trades on the ask (resp. bid) side. In
turn, positive (resp. negative)
mid-price variations imply an increase of the trading intensity on the bid (resp. ask) side.

Besides these important points, qualitative arguments showed that, provided $\Delta \lphi^T_0 \simeq 1$ and
$\Delta \lphi^N_0 < 0$, the ``long memory puzzle"  of the order flow
raised by Bouchaud et al. \cite{bouchaud1} can be addressed without any fine tuning of the model parameters: trades naturally appear as long-range
correlated over a wide range of lags while price variations are almost uncorrelated. Moreover, the same kind of arguments can explain the concave ("square-root" law) / relaxation typical market impact shape and
an almost vanishing permanent impact.

Let us first  point out that the model, as is, can be used as a ``stochastic price replayer" using as an input the true market order arrivals in place of the stochastic process $T_t$.
This allows one to ``replay" the price of a given historical period and, for instance, using it as an input price for any algorithm designed to estimate or manage a risk.

In this paper, we have presented the most basic form of the model.
It can be seen as a building block for more elaborate models depending on what it is meant to be used for.
There are many ways for extensions. Let us just go through some of them we have already developed or we are still working on. For instance, it would be important to have a model which not only takes into account the arrival times of the market orders but their volumes too.
This is a pretty easy extension since it can be done within the framework of marked Hawkes processes
for which straightforward extensions of all computations presented in this paper can be obtained. In the simplest form, one could use i.i.d. volumes $v_t$ for the market orders : at any time $t$ a market order arrives ($dT^+_t = 1$ or $dT^-_t = 1$) a volume $v_t$ is chosen randomly (using a given law). In its simplest form the new model consists in replacing the projection of \eqref{Hawkes4d} on the last two components by
\[
\lambda^N_t = \Phi^N \star dN_t + \Phi^I \star f(v_t)dT_t,
\]
where $f$ is a  function that describes how the volume impacts the price. It basically corresponds to what is generally referred into the literature by the ``instantaneous impact function".

In order to go further into the understanding of the underlying dynamics of the order-book, a very natural extension of the model, consists in using more dimensions in order to take into account limit/cancel orders. Thus, for instance, one way would be to introduce a new point process $L_t =
   \left(\begin{array}{c} L^-_t \\ L^+_t \end{array}\right)$ where $L^+_t$ (resp. $L^-_t$) is incremented by 1 whenever a limit order arrives at the best ask (resp. best bid) or a cancel order arrives at best bid (resp. best ask). One would then need to introduce the different kernels that account for the influence of $L$ on $T$ and $N$ and the kernels that account for the influence of $T$ and $N$ on $L$ itself.  The estimation procedure of the kernels can follow the exact same procedure described in Section \ref{sec:estimation}. Along the same line, i.e., by adding new dimensions to the 4d-Hawkes model presented in this paper, one could quantify the impact of a given exogenous news on the market order flow or directly on the price by estimating the corresponding kernel. Or, alternatively, one could consider a multi-agent models (e.g., adding 2 dimensions for each agent) and model/estimate the influence of a given agent on another one or on all the anonymous agents (as $\Theta^T$ does in \eqref{LHawkes4d}).
These extended framework would open the door to precise estimations and obvious interpretations in order to get better insights into to full order book dynamics.

\section{Annexes}
\subsection{Stability condition : proof of Proposition \ref{prop:stability}}
\label{proofstab}
In this section, we give the proof to the Proposition \ref{prop:stability}
For the process to be stable, we need the eigenvalues of the matrix $\lPhi_0$ to have a modulus smaller than 1 (cf hypothesis (H)).

\begin{lem} \label{lm1} {\bf (Eigenvalues of $\lPhi_0$)} \\
If $x$ is an eigenvalue of $\lPhi_0$, then it satisfies
\begin{equation}
\label{1}
(a^--x)(b^--x)-c^- = 0
\end{equation}
or
\begin{equation}
\label{2}
(a^+-x)(b^+-x)-c^+ = 0,
\end{equation}
where
$a^\pm = \lphitts_0\pm\lphittc_0$, $b^\pm = \lphinns_0\pm\lphinnc_0$ and $c^\pm = (\lphints_0\pm\lphintc_0)(\lphitns_0\pm\lphitnc_0)$.
(Let us point out that, using Notations \ref{not:delta}, $a^- = \Delta \lphitt_0$, $b^- = \Delta \lphinn_0$ and
$c^- = \Delta \lphint_0 \Delta \lphitn_0$).
\end{lem}
{\em Proof} \\
From \eqref{lphi} we get
\begin{equation}
det(\lPhi_0 -x\Id)=  det\left(\begin{array}{cc}
\lPhitt_0-x\Id & \lPhint_0 \\
\lPhitn_0 & \Id\lPhinn_0-x\Id
 \end{array} \right) = 0.
\end{equation}
Since all the matrices are bi-symmetric, they all commute and thus
\begin{equation}
det(\lPhi_0 -x\Id)=  det\left((\lPhitt_0-x\Id)(\lPhinn_0-x\Id)-\lPhint_0\lPhitn_0\right) = 0
\end{equation}
Moreover, they all diagonalize in the same basis and their eigenvalues are the sum and the difference between the self term and the cross term. Thus the eigein values of $(\lPhitt_0-x\Id)(\lPhinn_0-x\Id)-\lPhint_0\lPhitn_0$ satisfy either \eqref{1} or \eqref{2}. \\

\vskip .4cm
\noindent
We now have to study when the modulus of the roots of these equations  are smaller than 1.
\begin{lem} \label{lm2} {\bf (Condition for the roots to have a modulus smaller than 1)} \\
The roots of the equation
\[
x^2 -x(a+b) + ab-c = 0
\]
have a modulus smaller than 1 iff
\begin{itemize}
\item[(i)] either
\begin{equation}
\label{i}
c < -(a-b)^2/4 \mbox{~~~and~~~}  ab-c < 1
\end{equation}
\item[(ii)] or
\begin{equation}
\label{ii}
c > -(a-b)^2/4 \mbox{~~and~~~} |a+b| < \min(1+ab-c,2)
\end{equation}
\end{itemize}
\end{lem}
{\em Proof} \\
The discriminant of this second-order equation is $(a-b)^2+4c$, thus there are two cases
\begin{itemize}
\item[(i)] $c < -(a-b)^2/4$. In this case the roots are complex and conjugated one of the other. Thus their modulus is smaller than 1 iff their product is smaller than 1, i.e., $ab-c < 1$ which gives the result for (ii)
\item[(ii)] $c > -(a-b)^2/4$. In this case the roots are real. The condition for them to be smaller than 1 is
\[
1-(a+b)+ab-c > 0 \mbox{~~and~~} \frac{a+b} 2 < 1
\]
and the condition for them to be greater than -1 is
\[
1+(a+b)+ab-c > 0 \mbox{~~and~~} \frac{a+b} 2 > -1
\]
This synthesizes into case ii of Lemma
\end{itemize}

\vskip .4cm
\noindent
Using these two lemmas, let us study the stability condition corresponding to \eqref{1} and \eqref{2}.
Before starting, let us note that, since all the $\phi$ functions are positive functions (thus all the $\lphi_0$ are real positive), it is clear that
\begin{equation}
\label{pm}
|a^-| \le a^+,~~|b^-| \le b^+,~~|c^-| \le c^+.
\end{equation}

\vskip .2cm
\noindent
{\bf Proving \eqref{H} is a necessary condition for (H) to hold.}  \\
For this purpose, we just need to look at the condition for the roots of \eqref{2} to have a modulus smaller than 1.
Indeed, \eqref{2} falls in the case of  (ii) of Lemma \ref{lm2}, since $c = c^+ \ge 0$. Moreover \eqref{ii} is equivalent to
$|a^++b^+| < \min(1+a^+b^+-c^+,2)$ which is equivalent to
 \begin{equation}
 \label{11}
c^+ < (1-a^+)(1-b^+)  \mbox{~~and~~} a^+,~b^+ < 1
\end{equation}
which proves \eqref{H}

\vskip .4cm
\noindent
{\bf Proving \eqref{H} is a sufficient condition for (H) to hold. (and proving  \eqref{H1})}  \\
For that purpose, we suppose that \eqref{H} holds. We want to prove that the roots of both \eqref{1} and \eqref{2} have modulus strictly smaller than 1. We have just seen this is the case for the roots of \eqref{2}, we just need to check that it is also the case for the roots of  \eqref{1}.
For \eqref{2} the case (i) reads
\[
a^-b^--1 < c^- < -(a^--b^-)^2/4
\]
and the case (ii) (since $|a^- + b^-| \le  a^++ b^- <2$ according to \eqref{11})
 \[
-(a^--b^-)^2/4 <  c^- < 1+a^-b^--|a^-+b^-|
\]
Thus, merging these last two inequations, we get the following condition for the roots of \eqref{1}  to have a modulus strictly smaller than 1
 \begin{equation}
 \label{22}
a^-b^--1 < c^- < 1+a^-b^--|a^-+b^-|,
\end{equation}
which is nothing but \eqref{H1}.
Thus, in order to complete the proof of Proposition \ref{prop:stability}, we just need to prove that this last inequation holds (i.e., \eqref{H1} holds).

\noindent
Let  us first  notice that, since $|a^-| \le a^+ <1$ and $|b^-| \le b^+ <1$, one has
$2 a^-b^- \leq |a^-+b^-|$, and consequently
\[
1+a^-b^--|a^-+b^-| \le 1-a^-b^-.
\]
Since $1-a^- b^-  > 0$ the following inequation
\[
|c^-| < 1+a^-b^--|a^-+b^-|.
\]
is a sufficient condition for \eqref{22} to hold. Moreover since $|c^-| \le c^+$, using \eqref{H}, a sufficient condition for \eqref{22} to hold is
\[
(1-a^+)(1-b^+) \le 1+a^-b^--|a^-+b^-|.
\]
which is equivalent to
\begin{equation}
\label{33}
a^++b^+ - a^+b^+  \geq  |a^-+b^-| - a^-b^-.
\end{equation}
Since, one has
\begin{eqnarray}
  a^+ + b^+ -a^+ b^+ & = & a^+(1-b^+)+b^+
   \geq  |a^-|(1-b^+)+b^+ \\
  & \geq & b^+ (1-|a^-|)+|a^-|  \geq |b^-| (1-|a^-|)+|a^-| \\
  &\geq& |a^-|+|b^-| -|a^-||b^-|,
\end{eqnarray}
inequation \eqref{33} is implied by
\begin{equation}
\label{33bis}
|a^-|+|b^-| -|a^-||b^-|  \geq  |a^-+b^-| - a^-b^-.
\end{equation}
Thus, in order to complete the proof of Proposition \ref{prop:stability}, we just need to prove this last inequation.

\vskip .4cm
\noindent
In the case  $a^{-}$ and $b^{-}$ have the same sign, this inequation is a actually a strict equality, so it obviously holds.
Let us suppose that  $a^{-}$ and $b^{-}$ do not have the same sign. Without loss of generality, we can suppose that $a^- \le 0 \le b^-$.
We distinguish two cases :
\begin{itemize}
\item either $a^-\le -b^-\le 0$, in which case
\[
  |a^-+b^-| - a^-b^-  = -b^-(1+a^-)-a^- \le b^-(1+a^-)-a^- = |a^-| + |b^-| -|a^-||b^-|.
\]
\item or $-b^- < a^-<0$, in which case
\[
  |a^-+b^-| - a^-b^- = a^-(1-b^-)+b^- \le -a^-(1-b^-)+b^- = |a^-| + |b^-| -|a^-||b^-|.
\]
\end{itemize}

\subsection{Conditional expectation of a Hawkes process}
\label{annex_cond}
In this Section, we establish general results on $N$-dimensional Hawkes process $P = \{P^i\}_{1\le i \le N}$ and more particularly on the
expectations of $dP^i$ at time $t$ conditioned by the fact that the $P^j$ jumped at time 0. We mainly establish two results.
The fist one (Prop. \ref{th:cexpect}) links these conditional expectations with the auto-covariance function of $dP_t$ and, using previous results \cite{BaMuDaEPJB,BDHM2011}, allows us to derive an analytical formula as a function of the kernel. These results are used in the Section \ref{response} for characterizing the response function. The second one (Prop. \ref{th:fredholm})
proves that these expectations satisfy a Fredholm system that will be used in Section \ref{sec:estimation} for elaborating a general procedure for the kernel estimation.

The Hawkes process is defined by its kernel  $\Phi = \{\phi^i\}_{1\le i,j \le N}$ and the exogenous intensity $\mu = \{\mu^i\}_{1\le i \le N}$
through the equation
\begin{equation}
\label{lambda}
\lambda_t = \mu + \phi \star  dP_t,
\end{equation}
where $\lambda_t$ is the conditional intensity of $P$ at time $t$.
We consider that the process is stable, i.e., the eigenvalues of the matrix $\lPhi_0$ have modulus smaller than 1.
We set
\begin{equation}
\label{Lambda}
\Lambda = \E(\lambda_t) = \left( \Id -\lPhi_0\right)^{-1} \mu.
\end{equation}
Finally for all $t$ and $\forall ~i,j$ such that $1\le i,j \le N$, we define $g_t = \{g^{ij}_t \}_{1\le i,j \le N}$ with
\begin{equation}
\label{gij}
g^{ij}_t dt = \E(dP^i_t| dP^j_0 = 1) - \epsilon^{ij} \delta_t -\Lambda^i dt
\end{equation}
where $\E(dP^i_t| dP^j_0 = 1)$ is the conditional expectation of $dN^i_t$ knowing that $N^j_t$ jumps at $t=0$, $\delta_t$ is the dirac distribution and $\epsilon^{ij}$ is always 0 except for $i=j$ for which it is equal to 1.
Since $\E(dN^i_t| dN^i_0 = 1)$ is singular at $t=0$ we substracted this singular component.
In the following $g_t$ will refer to the matrix whose elements are the $g^{ij}_t$, i.e.,
\begin{equation}
\label{g}
 g_t = \{g^{ij}_t\}_{1\le i,j \le N}.
\end{equation}
We are ready to state our first result.
\begin{proposition}
\label{th:cexpect}
Let $C_{t,t+\tau}$ be the infinitesimal covariance matrix (without the singular part) as defined by
\[
C_{t,t+\tau} = \{C^{ij}_{t,t+\tau}\}_{1\le i,j \le N},
\]
with
\[
C^{ij}_{t,t+\tau} dt d\tau= Cov(dP^i_{t+\tau}, dP^j _{t}) -\epsilon^{ij}\Lambda^i\delta_{\tau} dt.
\]
Then $g_t$ and $C_{t,t+\tau}$ are linked through the relation
\begin{equation}
\label{gcov1}
g_{\tau} = C_{t,t+\tau}\Sigma^{-1}.
\end{equation}
Using the analytical formula proved in \cite{BaMuDaEPJB,BDHM2011} for $C^{ij}$,one gets
\begin{equation}
\label{gcov}
g_\tau = \Psi +\Sigma \tilde \Psi\Sigma^{-1} + \Psi \star \Sigma \tilde \Psi_\tau^\dagger\Sigma^{-1},
\end{equation}
where $\tilde \Psi_\tau = \Psi_{-\tau}$ and $\Psi$ is defined such that $(\Id-\lPhi_z)^{-1} = (\Id+\lPsi_z)$.
In the Fourier domain, this last equation corresponds to
\begin{equation}
\label{fgcov}
\lap g_z = (\Id-\lPhi_z)^{-1}\Sigma (\Id-\lPhi_z^\dagger)^{-1}\Sigma^{-1} -\Id.
\end{equation}
\end{proposition}
{\em Proof of the Proposition.} \\
This is a pretty straightforward computation :
\[
C^{ij}_{t,t+\tau}dt d\tau = \E(dP_{t+\tau}^idP_{t}^j) - \Lambda^i \Lambda^jdt d\tau -\epsilon^{ij}\Lambda^i\delta_\tau dt.
\]
which gives
\[
C^{ij}_{t,t+\tau}dt d\tau =  \E(dP_{t+\tau}^i| dP_{t}^j=1)Prob(dP_{t}^j=1) - \Lambda^i \Lambda^jdt d\tau -\epsilon^{ij}\Lambda^i\delta_\tau dt.
\]
Using the stationnarity and dividing by $dt$, one gets
\[
C^{ij}_{t,t+\tau}d\tau = \E(dP_{\tau}^i| dP_0^j=1)\Lambda^j - \Lambda^i \Lambda^jd\tau - \epsilon^{ij} \Lambda^i \delta_\tau.
\]
Then, using the definition \eqref{gij} of $g^{ij}$
\[
C^{ij}_{t,t+\tau} d\tau =   g^{ij}_{\tau} \Lambda^j d\tau
\]
which gives \eqref{gcov1}.
From \cite{BaMuDaEPJB,BDHM2011}, we know that
\[
C_{t,t+\tau}  = \Psi_\tau \Sigma + \Sigma\tilde \Psi_\tau + \Psi \star \Sigma \tilde \Psi_\tau^\dagger,
\]
This last equation along with \eqref{gcov1} leads to \eqref{gcov} and then to  \eqref{fgcov}.

\vskip .3cm
\noindent
The next result corresponds to the following proposition :
\begin{proposition}
\label{th:fredholm}
Using the notations above, the density $g_t$ satisfies the following fredholm system of integral  equations for positive $t$
\begin{equation}
\label{fredholm}
g_t = \phi \star (\delta\Id + g_t),~~~~\forall t > 0.
\end{equation}
Moreover,
for $t<0$ we have
\begin{equation}
\label{fredholm1}
g_t\Sigma = \Sigma g_{-t}^{\dagger}.
\end{equation}
\end{proposition}
\noindent
{\em Proof of the Proposition.} \\
\begin{itemize}
\item[$\bullet$] Proof of \eqref{fredholm}.
We consider $t > 0$
$$
g^{ij}_t = \E(dP^i_t| dP^j_0 = 1) -\Lambda^i = \E(\lambda^i_t| dP^j_0 = 1)-\Lambda^i
$$
Using \eqref{lambda},
$$
g^{ij}_t = \mu^{i}+ \sum_{k=1}^N \phi^{ik} \star \E(dP^k_t | dP^j_0 = 1)-\Lambda^i.
$$
Using \eqref{gij}, we get
$$
g^{ij}_t = \mu^{i}+ \sum_{k=1}^N \phi^{ik} \star (g_t^{kj} + \epsilon^{kj}\delta_t +\Lambda^k) -\Lambda^i.
$$
And consequently
$$
g^{ij}_t = \mu^{i}+  \sum_{k=1}^N ||\phi^{ik}|| \Lambda^k   -\Lambda^i + \phi^{ij}_t + \sum_{k=1}^N \phi^{ik} \star g^{kj}_t
$$
However, the vector formulation of $\mu^{i}+  \sum_{k=1}^N ||\phi^{ik}|| \Lambda^k   -\Lambda^i$ is nothing but
$\mu+  (\lphi_0-\Id)\Lambda$ which is 0 according to  \eqref{Lambda}, thus
$$
g^{ij}_t = \phi^{ij}_t + \sum_{k=1}^N \phi^{ik} \star g^{kj}_t
$$
which gives  \eqref{fredholm}.

Let us note that we could have derived directly the system \eqref{fredholm} from \eqref{fgcov}. Indeed, \eqref{fgcov}  gives
\[
(\Id-\lPhi_z) \lap g_z = \Sigma (\Id-\lPhi_z^\dagger)^{-1}\Sigma^{-1} -\Id + \lPhi_z.
\]
In the time domain, $\lPhi_z^\dagger$ is supported by $\mathds{R}^-$, thus going back to the time domain and restricting to $t >0$ directly leads to \eqref{fredholm}.

\item[$\bullet$]  Proof of \eqref{fredholm1}.
Let $t<0$.
$$
\E(dP^i_t | dP^j_0 = 1) = Prob(dP^i_t=1 | dP^j_0 = 1) = \frac { Prob(dP^i_t=1 , dP^j_0 = 1)}{Prob(dP^j_0 = 1)}
$$
Since $P$ is stable, $dP$ is stationary, $Prob(dP^j_0 = 1) = Prob(dP^j_t = 1) = \Lambda^j$, thus
$$
\Lambda^j\E(dP^i_t | dP^j_0 = 1) = Prob(dP^i_t=1 , dP^j_0 = 1) =
 \E(dP^j_{-t} | dP^i_0 = 1) \Lambda^i
$$
Consequently :
$$
\Lambda^jg^{ij}_t  = \Lambda^ig^{ji}_{-t}
$$
\end{itemize}

\section*{Acknowledgments}
The authors thank Sylvain Delattre, Marc Hoffmann, Charles-Albert Lehalle and Mathieu Rosenbaum for useful discussions.
We gratefully acknowledge financial
support of the chair {\it Financial Risks} of the {\it Risk Foundation} and of
the chair QuantValley/Risk Foundation: Quantitative Management Initiative.
The financial data used in this paper  have been  provided by the company {\em QuantHouse
EUROPE/ASIA}, http://www.quanthouse.com.


\end{document}